\documentclass[aps,prd,preprint,nofootinbib,amsmath,amssymb,showpacs,longbibliography,singlecolumn,superscriptaddress]{revtex4-1}

\usepackage{epsfig}
\usepackage{bm}
\usepackage{amssymb}
\usepackage{amsmath}
\usepackage{dsfont}
\usepackage{color}
\usepackage{subfigure}
\usepackage{dcolumn}
\usepackage{mathtools}%

\usepackage[utf8]{inputenc}

\usepackage{amsmath}
\usepackage{amssymb}
\usepackage{amsthm}
\usepackage{amscd, amsfonts, mathrsfs}
\usepackage{cases}
\usepackage{verbatim} 
\usepackage{epsf}
\usepackage[bookmarksnumbered,pdfpagelabels=true,plainpages=false,colorlinks=true,
            linkcolor=black,citecolor=black,urlcolor=black]{hyperref}

\theoremstyle{plain}

\theoremstyle{definition}

\allowdisplaybreaks

\newcommand*{\Scale}[2][4]{\scalebox{#1}{$#2$}}%
\newcommand{\forceds}[1]{\Scale[1.1]{\ensuremath{\mathrlap{{{#1}}}\hspace{1.4pt}#1}}}
\newcommand{\cvec}[1]{\ensuremath{\bm {#1}}}
\newcommand{\cmat}[1]{\ensuremath{\mathds{{#1}}}}
\newcommand{\cmatgreek}[1]{\ensuremath{\forceds{{#1}}}}

\newcommand{\be}{\begin{equation}}
\newcommand{\ee}{\end{equation}}
\newcommand{\bea}{\begin{eqnarray}}
\newcommand{\eea}{\end{eqnarray}}

\newcommand{\bml}{\begin{subequations}}
\newcommand{\eml}{\end{subequations}}

\newcommand{\bbm}{\begin{bmatrix}}
\newcommand{\ebm}{\end{bmatrix}}
\newcommand{\bvm}{\begin{vmatrix}}
\newcommand{\evm}{\end{vmatrix}}

\begin{document}

\title{Stochastic fluctuations in relativistic fluids: causality, stability, and the information current}
\date{\today}

\author{Nicki Mullins}
\email{nickim2@illinois.edu}
\affiliation{Illinois Center for Advanced Studies of the Universe\\ Department of Physics, 
University of Illinois at Urbana-Champaign, Urbana, IL 61801, USA}

\author{Mauricio Hippert}
\email{hippert@illinois.edu}
\affiliation{Illinois Center for Advanced Studies of the Universe\\ Department of Physics, 
University of Illinois at Urbana-Champaign, Urbana, IL 61801, USA}

\author{Jorge Noronha}
\email{jn0508@illinois.edu}
\affiliation{Illinois Center for Advanced Studies of the Universe\\ Department of Physics, University of Illinois at Urbana-Champaign, Urbana, IL 61801, USA}

\begin{abstract}
We develop a general formalism for introducing stochastic fluctuations around thermodynamic equilibrium which takes into account, for the first time, recent developments on the causality and stability properties of relativistic hydrodynamic theories. 
The method is valid for any covariantly stable theory of relativistic viscous fluid dynamics derived from a covariant maximum entropy principle. We illustrate the formalism with some applications, showing how it could be used to consistently introduce fluctuations in a model of relativistic heat diffusion, and in conformally invariant Israel-Stewart theory in a general hydrodynamic frame. The latter example is used to study the hydrodynamic frame dependence of the symmetric two-point function of fluctuations of the energy-momentum tensor.     
\end{abstract}

\maketitle

\tableofcontents

\section{Introduction}

Relativistic fluid dynamics \cite{Rezzolla_Zanotti_book} is an important tool in the description of vastly different physical systems, such as the quark-gluon plasma formed in ultrarelativistic heavy-ion collisions \cite{Romatschke:2017ejr} and  accretion disks surrounding supermassive black holes \cite{EventHorizonTelescope:2022urf}. Early models of relativistic hydrodynamics were constructed in the mid-twentieth century by Eckart \cite{PhysRev.58.919} and Landau and Lifshitz \cite{Landau1987Fluid}, but these were later found to possess unphysical behavior signaled by causality violation \cite{PichonViscous} and the fact that in such theories the global equilibrium state is not stable with respect to small disturbances in all Lorentz frames \cite{Hiscock:1985zz}.

These issues are not inherent to the formulation of viscous fluids in relativity. Recently, a general formulation of first-order relativistic hydrodynamics has been developed by Bemfica, Disconzi, Noronha, and Kovtun (BDNK) \cite{BemficaDisconziNoronha,Kovtun:2019hdm,Bemfica:2019knx,Hoult:2020eho,Bemfica:2020zjp} in which general conditions are given that ensure that the hydrodynamic equations of motion are causal and strongly hyperbolic \cite{Bemfica:2020zjp} in the fully nonlinear regime (including shear, bulk, and conductivity effects), and the equilibrium state is stable against small perturbations for all inertial frames. The latter feature, called \emph{covariant stability}, is interpreted as a dynamical property of the theory valid on-shell, i.e., along solutions of the field equations. This new first-order approach extends previous developments \cite{EckartViscous,LandauLifshitzFluids,Van:2007pw,Tsumura:2007wu,TempleViscous,TempleViscous2,TempleViscous3} by fully taking into account the fact there is no unique definition for the hydrodynamic variables (temperature, flow velocity, chemical potential) in an out of equilibrium state. Each definition of such variables is called a hydrodynamic frame\footnote{The word “frame” has also other different meanings in
relativity, e.g., inertial frames related by a Lorentz transformation, local rest frame, etc. However, the different uses of the word frame can be clearly distinguished depending on the context.} \cite{Kovtun:2012rj}, and the definitions used by Eckart, and Landau and Lifshitz are simple examples of (classes of) hydrodynamic frames. Thus, the non-equilibrium corrections that define the fluid's constitutive relations must allow for the presence of all the possible terms, compatible with the symmetries, involving first-order spacetime derivatives of the hydrodynamic variables which vanish in equilibrium. By exploring hydrodynamic frames different from Eckart and Landau-Lifshitz, BDNK showed that there is an infinite set of consistent definitions of the hydrodynamic variables out of equilibrium that ensures causal and stable evolution at first-order in derivatives.  

An earlier solution to the causality and stability problems involved the so-called second-order approaches, see \cite{1967ZPhy..198..329M,MIS-6,Baier:2007ix,Denicol:2012cn}. Following the work of Mueller \cite{1967ZPhy..198..329M}, and Israel and Stewart \cite{MIS-6}, causality and stability can be restored in the linear regime around equilibrium \cite{Hiscock_Lindblom_stability_1983,Olson:1989ey} using a qualitatively different idea than the one employed in the general first-order formalism mentioned above. In fact, in second-order theories, the viscous fluxes (such as the shear-stress tensor) obey additional equations of motion (derived using multiple approaches \cite{MIS-6,Baier:2007ix,Denicol:2012cn,MuellerRuggeriBook,JouetallBook}) that describe how these dissipative quantities evolve towards their first-order, universal behavior. A natural way to obtain such equations of motion  is to follow \cite{MIS-2} and employ a covariant maximum entropy principle using a suitably defined form for the entropy current out of equilibrium. Second-order hydrodynamic models are amply used in applications, especially when it comes to the hydrodynamic simulations of the quark-gluon plasma formed in ultrarelativistic heavy-ion collisions, see for instance, \cite{Romatschke:2017ejr}. 

A connection between first-order and second-order causal theories was recently discussed in \cite{Noronha:2021syv}. In that work, second-order hydrodynamic equations were obtained by taking into account all the possible deviations from equilibrium up to second order in a general hydrodynamic frame, using the covariant maximum entropy principle \cite{MIS-2}. This introduces new transient non-hydrodynamic degrees of freedom into the second-order theory that are not present in Eckart and Landau-Lifshitz frames. By carefully truncating this theory to first-order in derivatives, one recovers all the BDNK terms in the constitutive relations, showing how BDNK can emerge from second-order hydrodynamics formulated in a general hydrodynamic frame \cite{Noronha:2021syv}. In this case, both the second-order theory and its first-order limit can be causal and stable, though that requires hydrodynamic frames different than the Eckart and Landau-Lifshitz frames. 

Despite the important developments that occurred in recent years, many questions still remain concerning the formulation of relativistic hydrodynamics. For instance, it has been known for many years that there is a deep connection between causality and stability in relativistic fluids \cite{Hiscock_Lindblom_stability_1983,Pu:2009fj}. In the linear regime, causality is equivalent to stating that retarded Green's functions\footnote{In quantum field theory, causality imposes that the commutator between observables separated by spacelike intervals vanishes \cite{Peskin_book}.} vanish outside the future light cone \cite{Aharonov:1969vu}. Stability refers to the property that in fluids one expects that small disturbances around the equilibrium state remain bounded at arbitrarily large times. In a relativistic system, this property should be valid in all inertial frames.

In fact, Ref.\ \cite{Bemfica:2020zjp} proved a theorem that states that if a causal and strongly hyperbolic theory is stable in a given reference frame, it must be stable in any reference frame. This natural result follows from the fact that, in a causal relativistic theory, if a response function is analytical in the upper half of the complex frequency plane (i.e., a retarded Green's function) in a given Lorentz frame, no singularities can enter that region when using other reference frames. Furthermore, Ref.\ \cite{Gavassino:2020ubn} showed that theories in which the entropy is maximal in equilibrium, in a covariant way, are necessarily causal in the linear regime (such theories are also strongly hyperbolic, see \cite{Gavassino:2023odx}). A key result was later presented in \cite{Gavassino:2021owo} where it was shown that only in causal theories of relativistic fluid dynamics the stability properties of  disturbances around the global equilibrium are independent of the Lorentz frame. In fact, let us follow \cite{Gavassino:2021owo} and imagine that a spontaneous thermal fluctuation has occurred somewhere in the fluid according to some inertial observer A, which then sees this fluctuation dissipating away as a function of time. In an \emph{acausal} fluid dynamic theory, there is always an inertial observer B, connected to A via a Lorentz transformation, that will  disagree about the fate of the fluctuation, observing it to grow as a function of time (see \cite{Gavassino:2021owo}). This occurs because, in an acausal theory, the chronological sequence of events is not preserved by Lorentz transformations. Causality is then needed to make sure that every single possible inertial observer in a relativistic fluid agrees about the dissipation of spontaneous thermal fluctuations. Of course, this property is not present in studies where stochastic noise is included in the acausal viscous fluid dynamic theories derived by Landau and Lifshitz, and Eckart.  

The connection between causality and stability in relativistic systems was further strengthened by Refs.\ \cite{Heller:2022ejw,Gavassino:2023myj}. In fact,  \cite{Gavassino:2023myj} showed that covariant stability can be achieved by imposing that the dispersion relations of excitations obey the inequality $\mathfrak{Im} \, \omega(k) \leq |\mathfrak{Im} \, k|$, which was previously introduced in Ref.\ \cite{Heller:2022ejw} as a necessary condition for causality that leads to new bounds on transport coefficients describing stable phases of matter. 

The developments mentioned above focused only on the deterministic behavior associated with dissipative aspects of relativistic fluids, modeled via a set of nonlinear PDEs. However, a complete description of relativistic hydrodynamic phenomena also requires the inclusion of effects coming from the ubiquitous stochastic fluctuations that occur even in the equilibrium state \cite{landau_statistical_part_II}. There have been a number of works in the past years which investigated the interplay between dissipation and fluctuations in the formulation of relativistic hydrodynamics \cite{Calzetta:1997aj, Dunkel:2008ngc, Kapusta:2011gt, Kovtun:2012rj, Young:2013fka,Kumar:2013twa,Murase:2016rhl,Kapusta:2014dja,Akamatsu:2016llw,Akamatsu:2017rdu,Murase:2019cwc,An:2019osr,An:2020vri,An:2019csj, De:2020yyx, Torrieri:2020ezm, Dore:2021xqq, De:2022tkb, Abbasi:2022rum}.  
Applications to many problems of interest for heavy-ion collisions include \cite{Young:2014pka, Sakai:2017rfi, Singh:2018dpk, Sakai:2020pjw, Kuroki:2023ebq}, the dynamics of critical phenomena \cite{Stephanov:2017ghc, Nahrgang:2018afz, Martinez:2019bsn, Rajagopal:2019xwg, An:2019csj, Nahrgang:2020yxm, Dore:2020jye, Dore:2022qyz, Du:2020bxp, Pradeep:2022mkf,An:2022jgc}, hydrodynamic long-time tails \cite{Kovtun:2003vj, Kovtun:2011np,Akamatsu:2016llw,Akamatsu:2017rdu,Martinez:2017jjf, Martinez:2018wia}, and turbulence \cite{Calzetta:2020wzr}. Significant progress has also been achieved recently in the formulation of stochastic hydrodynamics using powerful field theory techniques \cite{Grozdanov:2013dba,Kovtun:2014hpa,Harder:2015nxa,Crossley:2015evo, Haehl:2015pja, Haehl:2015uoc, Haehl:2016pec, 
 Jensen:2017kzi,Glorioso:2017fpd,Liu:2018kfw,Chen-Lin:2018kfl,Jensen:2018hse, Haehl:2018lcu, Jain:2020vgc,Jain:2020zhu,Abbasi:2022aao}. 

The inclusion of stochastic noise is only sensible if the system under consideration is stable not only \emph{on-shell}, i.e. along solutions of the classical equations of motion, but also \emph{off-shell}, that is, against spontaneous fluctuations. Thus, a complete account of stochastic relativistic hydrodynamics must be grounded on hydrodynamic theories that are stable \emph{off-shell}, in a \emph{relativistic sense} \cite{Gavassino:2021kjm}.

In this work, we investigate the linear stochastic dynamics displayed by causal theories of relativistic fluids derived from a covariant maximum entropy principle. A new formalism is presented that can be used to determine the correlators of fluctuations in such theories using the thermodynamic information current introduced in \cite{Gavassino:2021kjm}, which determines the probability distribution for fluctuations around the equilibrium state. This provides a relativistic generalization of the well-known approach proposed by Fox and Uhlenbeck \cite{doi:10.1063/1.1693183} to describe non-relativistic hydrodynamic fluctuations. 
This new framework provides a simple way to determine correlation functions at spacelike separations that can be used to study fluctuating hydrodynamic theories in a general hydrodynamic frame, with the added benefit that covariant stability constraints are already built in. We apply our theory to some specific examples such as a simplified relativistic model of heat transport, and the case of conformally invariant (charge-neutral) Israel-Stewart theory in a general hydrodynamic frame \cite{Noronha:2021syv}. 
 
This paper is organized as follows. In section \ref{relativistic_thermodynamics} we review some basic properties of thermodynamic systems in relativity and explain the properties of the information current introduced in \cite{Gavassino:2020ubn}. We use the information current to explain the issues that appear when adding noise to acausal theories, such as charge-neutral conformal Landau-Lifshitz theory. Section \ref{relativistic_fluctuations} shows how to determine relativistic hydrodynamic fluctuations from the information current for a variety of systems. The relativistic generalization of Fox and Uhlenbeck's approach is presented in \ref{FoxUhlenbeck}.  Applications of our results appear in Sections \ref{Heat_equation_causal}  and \ref{gIS_fluctuations}. Our final remarks are presented in \ref{conclusions}. 

\emph{Notation}: We use natural units $\hbar=c=k_B=1$, a 4-dimensional Minkowski spacetime metric $g_{\mu\nu}$ with a mostly plus signature, and Greek indices
run from 0 to 3 while lower-case Latin indices run from 1 to 3 and upper-case Latin indices run across the space of thermodynamic variables. The four-momentum is written as $k^{\mu} = (\omega, k)$.

\section{Relativistic thermodynamics and the information current}
\label{relativistic_thermodynamics}

In this section, we first review the fundamental properties that relativistic fluids must have in equilibrium. We closely follow Ref.\ \cite{Gavassino:2021kjm} and consider a relativistic fluid in contact with a heat/particle  bath, and the system fluid plus bath is isolated. If the whole system evolves spontaneously from a state 1 to 2, standard thermodynamics \cite{landau_statistical_1980} dictates that the total entropy of the system should not decrease, $\Delta S_\mathrm{total} = \Delta S+\Delta S_\mathrm{bath} \geq 0$, where $\Delta S = S_{(2)} - S_{(1)}$ is the entropy difference between these states. Assuming that the heat bath is sufficiently large, one finds that 
\be
\Delta S_\mathrm{total} = \Delta(S+\alpha^*_I Q^I)\geq 0 ,
\ee
where $\alpha_I^* = \alpha_I^\textrm{bath}$ is the thermodynamic conjugate  associated with the conserved charges $Q^I$ of the system. Therefore, the functional $\Phi = S + \alpha_I^* Q^I$ is maximized in equilibrium \cite{Gavassino:2021kjm,landau_statistical_1980}.  
\begin{figure}[h]
    \centering    \includegraphics[width=0.40\textwidth]{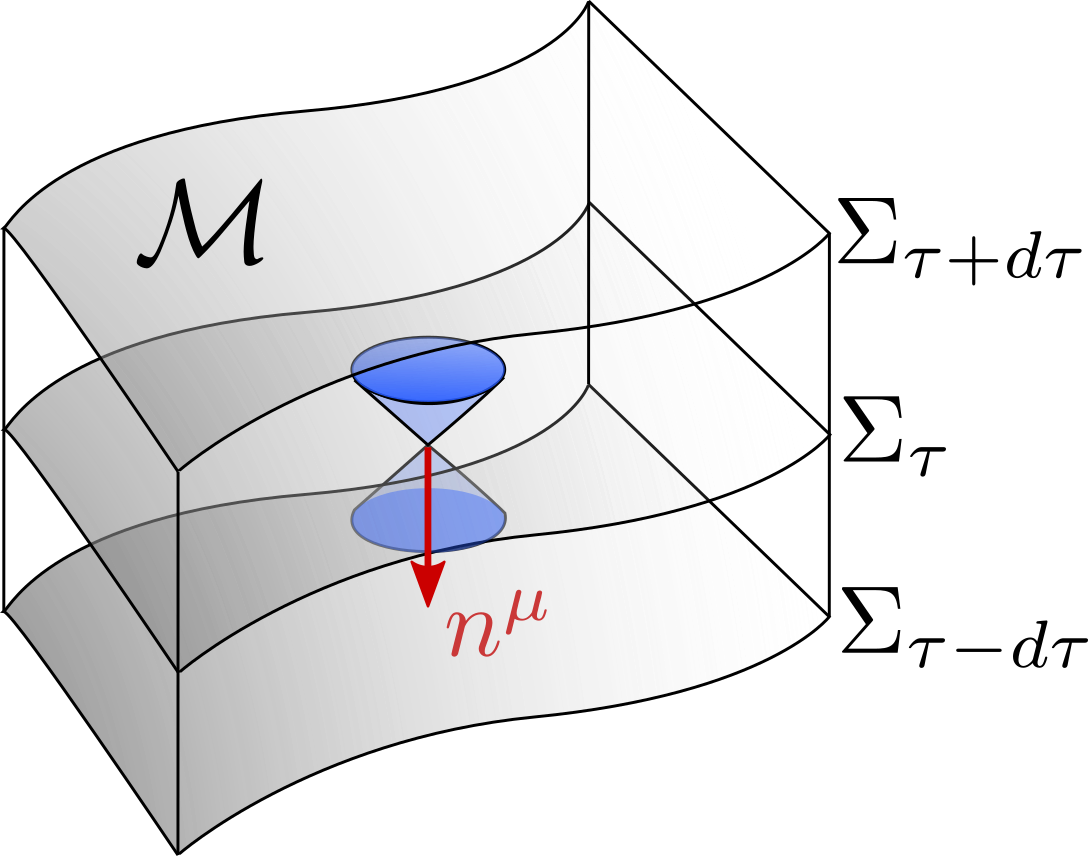}
    \caption{A foliation of spacetime, where $n^\mu$ is a past-directed, timelike, unit 4-vector normal to $\Sigma$.}
    \label{fig:foliation}
\end{figure}
Hence, for an arbitrary three-dimensional Cauchy surface $\Sigma$ (see Fig.\ \ref{fig:foliation}), one must impose that 
\be
E[\Sigma] = -\delta \Phi [\Sigma] = \int_\Sigma d\Sigma\, E^\nu n_\nu \geq 0 ,
\ee
where
\be
E^\mu = -\delta s^\mu - \alpha_I^* \delta J^{I\mu},
\ee
and $n^{\mu}$ is a timelike, past-directed ($n^0<0$) unit 4-vector  normal to $\Sigma$, $s^\mu$ is the entropy current of the fluid, $J^{I\mu}$ are the different conserved currents associated with the charges $Q^I$, and ``$\delta$" is an arbitrary finite perturbation of the equilibrium state. 

Consider the case where the conserved currents are the energy-momentum tensor and a 4-current associated with some global symmetry and a chemical potential $\mu$. In this case, one finds \cite{Gavassino:2021kjm}
\begin{equation}\label{eq:defE}
    E^{\nu}  = - \delta s^{\nu} - \frac{u_\lambda}{T} \delta T^{\lambda\nu} - \frac{\mu}{T} \delta J^{\nu} , 
    \quad \text{with} \quad \frac{\delta\Omega}{T} = E=\int_\Sigma d\Sigma\, n_\mu E^\mu ,
\end{equation}
so that $E T$ gives the variation of the grand thermodynamic potential $\delta \Omega$ \cite{Gavassino:2021kjm}. 

In order to have a consistent covariant description of such spontaneous thermodynamic processes, the current $E^\mu$ must obey a few properties. These are the properties necessary for the second law of thermodynamics to hold, complemented by the condition that no other state is as entropic as equilibrium, given a set of constraints.  They are:
\begin{enumerate}
    \item $E^{\mu} n_{\mu} \geq 0$ 
    for any past-directed, timelike unit vector $n_{\mu}$.
    \item $E^{\mu} n_{\mu} = 0$ \emph{if and only if} the perturbation of each hydrodynamic variable, $\delta \cvec{\phi}$, is equal to zero. 
    \item $\partial_{\mu} E^{\mu} \leq 0$.
\end{enumerate}
These conditions turn out to be precisely the ones for $\delta \Omega/T$ to behave as a Lyapunov functional, which leads to thermodynamic stability, fulfilling the  Gibbs stability criterion \cite{Gavassino:2021cli}. The conditions above also guarantee that the theory is causal in the linear regime, as shown in \cite{Gavassino:2021kjm}. It is crucial that these conditions hold for any timelike past-directed unit vector $n^\mu$ so that these are properties of $E^\mu$ alone.

At this point, it is important to remark here the difference between thermodynamic stability, in the sense defined above, and hydrodynamic stability \cite{Gavassino:2021kjm}. The former is used here to mean stability according to the covariant Gibbs stability criterion \cite{Gavassino:2021cli}, which generalizes the standard notion of thermodynamic stability \cite{landau_statistical_1980} to relativistic systems and must be valid also off-shell. Hydrodynamic stability is a dynamic property of the fluid equations of motion (hence, on shell) implying that perturbations around the equilibrium state are stable.

The properties listed above each have simple interpretations. The first property implies that no state is more probable than equilibrium, the second states that this equilibrium state is unique, and the final is an observer-independent statement of the second law of thermodynamics. Thus, these properties should be satisfied from purely thermodynamic arguments. Also, we note that $E^\mu$ in \eqref{eq:defE} near equilibrium is necessarily second-order in variations, given that at first-order it reduces to the standard covariant Gibbs relation \cite{MIS-6}
\be
\delta s^\nu = - \frac{u_\lambda}{T}\delta  T^{\lambda \nu} - \frac{\mu}{T}\delta J^\nu. 
\ee

The current $E^\mu$ can be interpreted as a current of information, in the following sense \cite{Gavassino:2021kjm}. The total $E$ is the net information carried by the perturbation and the fact that $E\geq 0$ (the Gibbs stability criterion \cite{Gavassino:2021cli}) implies that any perturbation increases our knowledge about the state (note that the equilibrium state is the one where $S_\mathrm{total}$ is maximal). The connection between $E^\mu$ and causality follows because condition (1) is equivalent to having $E^\mu$ being timelike or lightlike future-directed, $E^\mu E_\mu \leq 0$, $E^0\geq 0$, which in turn implies that the perturbations cannot transport information faster than light and information cannot leave the light cone, see \cite{Gavassino:2021kjm} for a proof.

To use the above quantities to derive a theory of relativistic thermodynamic fluctuations, the probability distribution of fluctuations is derived in terms of the information current found above. From standard thermodynamics \cite{landau_statistical_1980}, the probability of a given fluctuation in a closed system  is proportional to
\begin{equation} \label{thermal_dist_closed}
    w[\delta \cvec{\phi}] \propto e^{-\delta 
 \Omega/T},
\end{equation}
where $\delta \cvec{\phi}$ is a vector quantifying the perturbations in the space of hydrodynamic variables. This probability distribution is maximized when the free energy difference is minimized, just as expected of a system with an equilibrium state. Given that $\delta \Omega = E T$, we find that the probability distribution of fluctuations is determined by the information current
\begin{equation} \label{thermo_noise_probability}
    w[\delta \cvec{\phi}] \propto e^{- \int_{\Sigma}\, d\Sigma\, n_{\mu} E^{\mu}} .
\end{equation}
Indeed, one can think of this distribution as coming from a maximization of the entropy subject to a number of constraints corresponding to the conserved charges present in the system. Including simply the energy constraint then corresponds to the canonical ensemble, while including a conserved charge in addition to energy corresponds to the grand canonical ensemble.

{We also note that, because we are free to choose any timelike $n^\mu$ and any foliation in spacelike hypersurfaces, Eq.\ \eqref{thermo_noise_probability} can be used to calculate correlations between any set of points at spacelike separations from one another. That is,  correlations between any causally disconnected set of points follow from the information current in a straightforward manner. This should be contrasted with the non-relativistic theory, where analogous results are only sufficient to calculate equal-time correlators.

\subsection{Instability and acausality of Landau-Lifshitz theory: an information current perspective}
\label{navier_stokes_fluctuations}

In this section, we discuss how the information current can be used to understand the issues that appear in relativistic Navier-Stokes theory in the Landau-Lifshitz hydrodynamic frame \cite{LandauLifshitzFluids}. For simplicity, we neglect the effects coming from a conserved current and consider the case of a conformal fluid \cite{Baier:2007ix}. In the Landau-Lifshitz frame, the fluid four-velocity follows the flow of energy \cite{LandauLifshitzFluids}, i.e. 
\begin{equation}
    u_{\nu} T^{\nu\mu} = -\epsilon u^{\mu},
\end{equation}
where $T_{\mu\nu}$ is the energy-momentum tensor, $\epsilon$ is the energy density seen by an observer comoving with the fluid, which is defined to be the same energy density for a system in equilibrium (i.e., there are no out-of-equilibrium corrections to this quantity). With this constraint, the energy-momentum tensor of the conformal fluid at first-order in derivatives takes the form
\begin{equation}
    T^{\mu\nu} = \epsilon \left( u^{\mu} u^{\nu} + \frac{1}{3} \Delta^{\mu\nu} \right) -2\eta \sigma^{\mu\nu},
\end{equation}
where we used that the pressure $P=\epsilon/3$, $\eta$ is the shear viscosity coefficient, and $\sigma^{\mu\nu} = \Delta^{\mu\nu\alpha\beta}\partial_\alpha u_\beta$, with $\Delta^{\mu\nu\alpha\beta} = \frac{1}{2} \left( \Delta^{\mu\alpha} \Delta^{\nu\beta} + \Delta^{\mu\beta} \Delta^{\nu\alpha} \right) - \frac{1}{3} \Delta^{\mu\nu} \Delta^{\alpha\beta}$.

In the absence of other conserved currents, the equations of motion stem from energy-momentum conservation
\begin{equation}
    \partial_{\mu} T^{\mu\nu} = 0 .
\end{equation}
The corresponding entropy current is given by \cite{MIS-6}
\begin{equation}
    s^{\mu} = s u^{\mu} ,
\end{equation}
where $s = 4\epsilon / 3T$ is the equilibrium entropy density \cite{LandauLifshitzFluids}. Using the equations of motion, one can show that 
\be
T\,\partial_\mu s^\mu = 2\eta\,\sigma_{\mu\nu}\sigma^{\mu\nu},
\ee
which is positive semi-definite when $\eta>0$, even for arbitrarily large derivatives beyond the regime of applicability of first-order hydrodynamics. 
The above provides the starting point to determine how Landau-Lifshitz hydrodynamics behaves in the absence of noise. We will see below how the well-known issues with causality and instability \cite{Hiscock_Lindblom_stability_1983} that appear in this theory emerge from the point of view of the information current defined in the previous section. 

The information current for relativistic Navier-Stokes theory in the Landau-Lifshitz hydrodynamic frame can be obtained from \eqref{eq:defE}, and it reads
 \begin{equation} \label{E_LandauNS}
    E^{\mu}_{\mathrm{Landau}} = \frac{u^{\mu}}{8 \epsilon T} \delta \epsilon^2 + \frac{2 \epsilon u^{\mu}}{3T} \delta u^{\nu} \delta u_{\nu} + \frac{\delta \epsilon \delta u^{\mu}}{3 T} - \frac{2 \eta \delta u_{\nu}}{T} \Delta^{\mu\nu}_{\alpha\beta} \partial^{\alpha} \delta u^{\beta}, 
\end{equation}
where $\delta \epsilon$ and $\delta u^\mu$ are the deviations of these quantities with respect to the equilibrium state and $T$ and $u^\mu$ are equilibrium quantities.
We remind the reader that for thermodynamic stability in the sense of the Gibbs stability criterion to hold, $E^{\mu} n_{\mu} = 0$ if and only if $\delta \cvec{\phi} = 0$, for any timelike past directed unit vector $n^{\mu}$. However, projecting along some arbitrary $n^{\mu}$ of this kind, and setting the right-hand side of Eq.\ \eqref{E_LandauNS} to zero leads to a differential equation for $\delta u^\mu$, which has nontrivial solutions with $\delta u^{\mu} \neq 0$. 
Therefore, the requirements for the Gibbs stability criterion are not fulfilled by this theory. 
Such features will be present in any first-order theory in which the out-of-equilibrium degrees of freedom are written in terms of derivatives of equilibrium hydrodynamic variables.

One can see that this system will be  unstable for inertial observers with nonzero velocity, without using the explicit form of $\delta \sigma^{\mu\nu}$. This is simply because there is no term in the information current that goes with $\delta \sigma_{\mu\nu}\delta \sigma^{\mu\nu}$, so it is impossible to write this  as a perfect square ensuring that $n_{\mu} E^{\mu}$ is positive semi-definite for arbitrary $n^{\mu}$. This feature is inherent to first-order theories since a term that goes with $\delta \sigma^2$ will be second-order in derivatives\footnote{We thank L.~Gavassino for pointing this out to us.}. In second-order Israel-Stewart theories \cite{MIS-6} on the other hand, the entropy current is written generically in terms of out-of-equilibrium degrees-of-freedom up to second order (e.g. $\pi_{\mu\nu}\pi^{\mu\nu}$), guaranteeing that there are terms that depend on the square of out-of-equilibrium degrees-of-freedom in the information current. This feature is essential for constructing covariantly stable theories that obey the covariant Gibbs stability criterion in relativity. 

To see more specifically how fluctuations behave badly in relativistic Navier-Stokes theory, consider for simplicity the information current with $\delta \epsilon = 0$, 
\begin{equation} \label{E_landau_noeps}
    E_{\mathrm{Landau}}^{\mu} (\delta \epsilon = 0) = \frac{2 \epsilon u^{\mu}}{3T} \delta u^{\nu} \delta u_{\nu} - \frac{2 \eta \delta u_{\nu}}{T} \Delta^{\mu\nu}_{\alpha\beta} \partial^{\alpha} \delta u^{\beta} .
\end{equation}
We then want to contract this with an arbitrary past-directed, timelike unit vector $n^{\mu}$. It should be the case that $n_{\mu} E^{\mu}_{\mathrm{Landau}}\geq 0$, where the equality should hold only if $\delta u^{\mu} = 0$. One of these will be violated whenever 
\begin{equation}
    \frac{\eta}{T} n_{\mu} \delta u_{\nu} \Delta^{\mu\nu}_{\alpha\beta} \partial^{\alpha} \delta u^{\beta} \geq \frac{ \epsilon}{3T} n_{\mu} u^{\mu} \delta u^2 .
    \label{bad_NS}
\end{equation}
When these terms are equal, the out-of-equilibrium state will be equally probable as equilibrium but will have $\delta u^{\mu}$ and $\partial_{\mu} \delta u^{\nu}$ nonzero. To make matters worse, as the left-hand side becomes greater, the probability of the fluctuation occurring will increase according to Eq.\ \eqref{thermo_noise_probability}. This means that states with large $\partial_{\mu} \delta u^{\nu}$ will dominate the probability distribution, violating the basic assumption that equilibrium is the most probable state. One may argue that the offending terms with derivatives can be discarded since they are outside the regime of validity of the theory, but this would require removing the entire shear tensor $\delta \sigma^{\mu\nu}$ leaving only a perfect fluid (which, of course, does not exhibit fluctuations). 

This can be further understood by introducing the Knudsen number \cite{ChapmanCowling,Rischke_Denicol_book}, 
\begin{equation}
    \mathrm{Kn} = \frac{\ell}{L} ,
\end{equation}
where $\ell$ is the microscopic length scale (for example, the mean free path in a gas) and $L$ is the macroscopic scale associated with the gradients of macroscopic quantities. One may estimate the microscopic scale as 
\begin{equation}
    \ell =\frac{\eta}{\epsilon + P} = \frac{3\eta}{4\epsilon} ,
\end{equation}
while the macroscopic scale can be estimated from
\begin{equation}
    \frac{1}{L} = \frac{\sqrt{\delta u_{\nu} \delta \sigma^{\nu\rho} \delta u_\mu \delta \sigma^\mu_\rho}}{\delta u_\lambda \delta u^\lambda}.
\end{equation}
We then see that Eq.\ \eqref{bad_NS} occurs when 
\begin{equation}
    \mathrm{Kn} \gtrsim 1 ,
\end{equation}
meaning that $n_{\mu} E^{\mu} \leq 0$ when the Knudsen number is large. Nevertheless, since the probability of a given fluctuation goes with $\exp\{-E\}$, fluctuations for which $E < 0$ are exponentially favored. This means that the most probable fluctuations in Landau-Lifshitz theory, when observed in a general reference frame, will be those that violate the very assumptions behind its construction. This type of ``runaway" behavior for the fluctuations is consistent with the findings of \cite{Gavassino:2020ubn}, where it was shown that the total entropy of Landau-Lifshitz theory grows without bound, and the presence of unstable modes in this theory \cite{Hiscock:1985zz} represents the directions of growth of the entropy in the space of dynamically accessible states.

One could ask why these issues were not appreciated in the past in investigations about  thermal fluctuations in Landau-Lifshitz theory. The reason for this is subtle, though it is directly related to the fact that the stability of the equilibrium state must be a Lorentz invariant concept, valid for any observer and not just for the one comoving with the fluid. In fact, consider this theory and its fluctuations but now assume the local rest frame, which here corresponds to taking $n^{\mu} = -u^{\mu}$. In this limit, we find that 
\begin{equation}
    E_{\mathrm{Landau}} = -u_{\mu} E^{\mu}_{\mathrm{Landau}} = \frac{1}{8 \epsilon T} \delta \epsilon^2 + \frac{2\epsilon}{3T} \delta u^{\nu} \delta u_{\nu} + \frac{2\eta \delta u_{\nu}}{T} u_{\mu} \Delta^{\mu\nu}_{\alpha\beta} \partial^{\alpha} \delta u^{\beta}.
\end{equation}
Since $u_{\mu} \Delta^{\mu\nu}_{\alpha\beta} = 0$, the last term vanishes and we are left with only 
\begin{equation}
    E_{\mathrm{Landau}} = \frac{1}{8 \epsilon T} \delta \epsilon^2 + \frac{2 \epsilon}{3T} \delta u^{\nu} \delta u_{\nu} ,
\end{equation}
which matches the result for an inviscid fluid. Now, we see that this quantity is zero only when $\delta \cvec{\phi} = 0$, and otherwise positive, as would be expected of a thermodynamically stable system. Taking the local rest frame thus hides the issues present in Landau-Lifshitz first-order theory, but it does not remove them. In order to have a sensible response to thermodynamic perturbations in relativity, valid for any inertial frame, $E$ would have to be well-behaved for all $n^\mu$, not just the one given by the local rest frame. Otherwise, it is always possible to find inertial frames that do not see exponential decay but, rather, exponentially growing modes \cite{Gavassino:2021owo}. Therefore one concludes that Landau-Lifshitz theory is not stable against off-shell perturbations in a relativistic sense.

This result should not be a surprise as it is known that relativistic Navier-Stokes in the Landau frame is acausal \cite{Hiscock:1985zz}. As mentioned before, acausal dissipative theories cannot be stable in a Lorentz invariant manner \cite{Gavassino:2021owo}. This means that when considering stochastic fluctuations in relativistic theories, causality is a necessary condition for such spontaneous fluctuations to decay toward equilibrium for all inertial observers.

\section{Fluctuating relativistic fluid dynamics: general formalism}
\label{relativistic_fluctuations}

In this section, we use the thermodynamic probability distribution obtained from the information current to determine the noise correlators in a relativistic fluid. The main idea will follow the seminal paper by Fox and Uhlenbeck \cite{doi:10.1063/1.1693183}, which may be summarized as follows. A set of stochastic dynamical equations of motion are specified in linear order in deviations from equilibrium. Using these equations of motion, the conditional probability distribution for a state $\delta \cvec{\phi}$ (representing the hydrodynamic variables) is constructed, which measures the probability of the system being in a given state at time $t$, assuming some set of initial conditions. As $t \rightarrow \infty$, this conditional probability distribution converges to the thermodynamic probability distribution of Eq.\ \eqref{thermo_noise_probability}, which allows the correlator of any stochastic terms to be extracted. In this section, we will show how this can be used to determine correlators in a consistent way in covariantly stable and causal theories that stem from a maximum entropy principle.

Consider a set of linearized relativistic equations of motion given by \cite{Geroch:1990bw, Gavassino:2022roi}
\begin{equation} \label{stochastic_EoM}
    \cmat{M}^{\mu} \partial_{\mu} \delta \cvec{\phi} + \cmat{V} \delta \cvec{\phi} = \cvec{\Xi} ,
\end{equation}
where $\partial_\mu$ is the spacetime derivative, and we denote vectors in the space of fluctuating hydrodynamic variables with bold (eg. $\cvec{\phi}$), matrices in the space of hydrodynamic variables with script (eg. $\cmat{E}$), and $\cvec{\Xi}$ is a stochastic vector. Such a dynamical system will have a quadratic information current in $\delta \cvec{\phi}$ \cite{Gavassino:2021cli,Gavassino:2021kjm,Gavassino:2023odx}, which motivates the following definition for the information current
\begin{equation}
    E^{\mu} \equiv \frac{1}{2} \delta \cvec{\phi}^T\, \cmat{E}^{\mu}\, \delta \cvec{\phi} .
\end{equation}
A formal solution of Eq.\ \eqref{stochastic_EoM} can be obtained from the retarded Green's function $\cmat{G}_R(x,x')$ defined by 
\begin{equation}
    \left( \cmat{M}^{\mu} \partial_{\mu} + \cmat{V} \right) \cmat{G}_R(x,x') = \delta (x-x') \cmat{I} ,
\end{equation}
where $\cmat{I}$ is the identity matrix, and $\cmat{G}_R(x, x')$ satisfies 
\begin{equation}
    \cmat{G}_R(x,x') = 0 \:\: \mathrm{if} \:\: n_{\mu} (x^{\mu} - x'^{\mu}) \leq 0
\end{equation}
for all past-directed timelike $n^{\mu}$. Notice that a retarded Green's function with this property, valid for all inertial observers, is only possible in a causal theory \cite{Gavassino:2021owo}. The solution to Eq.\ \eqref{stochastic_EoM} is then given by 
\begin{equation}
    \delta \cvec{\phi} (x) = \delta \cvec{\phi}_h(x) + \int d^4x' \,\cmat{G}_R(x,x')\, \cvec{\Xi}(x') .
\end{equation}
$\delta \cvec{\phi}_h(x)$ is the homogeneous solution defined to satisfy the boundary conditions $\cvec{\phi}_0(x)$ on some initial spacelike hypersurface $\Sigma_0$. 

The conditional probability distribution for the system to be in state $\cvec{\phi}_f(x \in \Sigma_f)$ given the initial condition $\cvec{\phi}_0(x \in \Sigma_0)$ is given by 
\begin{equation} \label{conditional_P1}
\begin{split}
    \mathcal{P}_2^{\Sigma_f}[\delta \cvec{\phi}_f] & = \mathcal{P}[\delta \cvec{\phi}_f(x \in \Sigma_f)| \delta \cvec{\phi}_0 (x \in \Sigma_0)] \\
    & = \left\langle \delta \left[ \delta \cvec{\phi}_f(x) - \delta \cvec{\phi}_h(x) - \int d^4x'\, \cmat{G}_R(x,x') \, \cvec{\Xi}(x') \right] \right\rangle \\
    & = \left\langle \int \mathcal{D} \cvec{\lambda} \,\exp \left[ i \int_{\Sigma_f} d\Sigma_f \, \cvec{\lambda}(x) \left\{ \delta \cvec{\phi}_f(x) - \delta \cvec{\phi}_h(x) - \int d^4x' \cmat{G}_R(x,x')\, \cvec{\Xi}(x') \right\} \right] \right\rangle ,
\end{split}
\end{equation}
where $\cvec{\lambda}$ is some auxiliary variable. The expectation value is taken over noise realizations with probability
\begin{equation}
    P[\Xi] = \mathcal{N} \exp \left[ -\frac{1}{4} \int d^4x\, \cvec{\Xi}^T(x) 
    \cmat{Q}^{-1} \cvec{\Xi}(x) \right] 
\end{equation}
for some matrix $\cmat{Q}$ that gives the two-point correlation of $\cvec{\Xi}$:
\begin{equation}
    \langle \cvec{\Xi}(x) \cvec{\Xi}(x')
     \rangle = 2 \cmat{Q}\,\delta^{(4)}(x-x')
\end{equation}
and a normalization $\mathcal{N}$, such that 
\begin{equation}
    \langle \mathcal{O} \rangle = \int \mathcal{D} \Xi\, \mathcal{O}[\Xi]\, P[\Xi] .
\end{equation}
The path integral of Eq.\ \eqref{conditional_P1} can be evaluated by completing the square, in which case the conditional probability distribution becomes
\begin{equation}
    \begin{split}
        P_2^{\Sigma_f} & [\delta \phi_f] = \mathcal{N} \int \mathcal{D} \lambda \,\mathcal{D} \Xi'\, \exp \left( - \frac{1}{4}\int d^4x\, \cvec{\Xi}'^T(x) \cmat{Q}^{-1} \cvec{\Xi}'(x) \right) \exp \Bigg[ - \int_{\Sigma_f} d\Sigma_1 \cvec{\lambda}(x_1) \times \\
        & \times \int d^4x'\, \cmat{G}^{\dagger}(x_1,x') \, \cmat{Q} \int_{\Sigma_f} d\Sigma_2 \, \cmat{G}(x', x_2) \,\cvec{\lambda}(x_2) \Bigg] \exp \left[ i \int_{\Sigma_f} d\Sigma \, \cvec{\lambda}(x)\, (\delta \cvec{\phi}_f(x) - \delta \cvec{\phi}_h(x)) \right] ,
    \end{split}
\end{equation}
where $\cmat{Q}$ is inverted only in the subspace with nonzero eigenvalues, and $\cvec{\Xi}$ is a vector within this subspace. Then
\begin{equation}
    \cvec{\Xi}'(x) \equiv \cvec{\Xi}(x) + 2i \, \cmat{Q}\int_{\Sigma} \, d\Sigma \, \cmat{G}_R^{\dagger}(x,x')\, \cvec{\lambda}(x') ,
\end{equation}
where $(\cdots)^\dagger$ denotes taking a transpose and changing the order of space-time variables $x$ and $x'$. 
The path integral over $\Xi'$ (multiplied by $\mathcal{N}$) is just $\langle 1 \rangle = 1$. It is then convenient to define 
\begin{equation}\label{eq:RSigma}
    \cmat{R}_{\Sigma_f}(x_1, x_2) = 2 \int_{\Sigma_0}^{\Sigma_f} d^4x'\, \cmat{G}_R(x_1,x')\, \cmat{Q} \, \cmat{G}_R^{\dagger}(x',x_2) ,
\end{equation}
such that after completing the square once more, the conditional probability distribution is given by 
\begin{equation}
    P_2^{\Sigma_f}[\delta \phi_f] = \Tilde{\mathcal{N}} \exp \left[ -\frac{1}{2} \int_{\Sigma_f} d\Sigma_1\, d\Sigma_2\, [\delta \cvec{\phi}_f(x_1) - \delta \cvec{\phi}_H(x_1)]^T\, \cmat{R}^{-1}_{\Sigma_f}(x_1,x_2) \,[\delta \cvec{\phi}_f(x_2) - \delta \cvec{\phi}_H(x_2)] \right] .
\end{equation}
The term $\cmat{R}_{\Sigma_f}(x_1,x_2)$ is then a correlation for fluctuations of the thermodynamic variables $\cvec{\phi}$ around the homogeneous solution $\delta \cvec{\phi}_H$. A diagrammatic representation of this term is shown in Fig.\ \ref{fig:R_Sigma}. There, $x_1$ and $x_2$ are points on the final hypersurface $\Sigma_f$, which are both causally connected to some past spacetime point $x'$. A fluctuation at $x'$ would then propagate to both $x_1, x_2$ leading to some correlation at this point. Summing over correlations from all $x'$ between $\Sigma_0$ and $\Sigma_f$, yields $\cmat{R}_{\Sigma_f}(x_1,x_2)$. 

\begin{figure}[h!]
    \centering
    \includegraphics[width=0.25\textwidth]{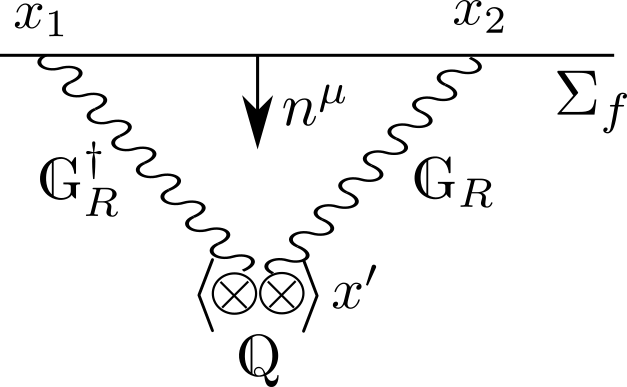}
    \caption{Diagrammatic representation of the two-point function $\cmat{R}_{\Sigma_f}(x_1,x_2)$ in Eq.\ \eqref{eq:RSigma}, where $x_1,x_2 \in \Sigma$. Local correlations of the  stochastic noise in $x'$, represented by the $\otimes$ symbols,  propagate to both points with the retarded Green's function $\cmat{G}_R$, represented by the squiggly lines, making the two points correlated. Contributions are summed over all points $x'$ between the initial and final hypersurfaces $\Sigma_0$ and $\Sigma_f$.  }
    \label{fig:R_Sigma}
\end{figure}

At long times, the homogeneous solution $\delta \cvec{\phi}_H$ is damped away and the conditional probability distribution becomes independent of the initial conditions
\begin{equation}
    P_1[\delta \phi_f] = \lim_{t \rightarrow \infty} P_2[\delta \phi_f | \delta \phi_0] .
\end{equation}
This probability distribution can then be identified with the probability distribution in thermodynamic equilibrium 
\begin{equation}
    w[\delta \phi] \propto e^{-\int_{\Sigma} d\Sigma \, n_{\mu} E^{\mu}} ,
\end{equation}
determined in Eq.\ \eqref{thermo_noise_probability}. If follows that for all $\delta \cvec{\phi}(x)$, 
\begin{equation}
    \begin{split}
     \lim_{(\tau_f - \tau_0) \rightarrow \infty}     \int_{\Sigma} d\Sigma_1\, d\Sigma_2\, \delta \cvec{\phi}(x_1) \cmat{R}^{-1}_{\Sigma_f}(x_1,x_2) \delta \cvec{\phi}(x_2) & = \int_{\Sigma} d\Sigma \, n_{\mu} \delta \cvec{\phi}(x) \cmat{E}^{\mu} \delta \cvec{\phi}(x) .
    \end{split}
\end{equation}
 This implies that 
\begin{equation}
  \lim_{(\tau_f - \tau_0) \rightarrow \infty}  \cmat{R}^{-1}_{\Sigma_f}(x,x') = n_{\mu} \cmat{E}^{\mu} \delta^{(3)}(x - x') ,
\end{equation}
or 
\begin{equation} \label{R=nE}
  R_{\infty} \equiv \lim_{(\tau_f - \tau_0) \rightarrow \infty}  \cmat{R}_{\Sigma_f}(x, x') = (n_{\mu} \cmat{E}^{\mu})^{-1} \delta^{(3)}(x-x'),
\end{equation}
where $\delta^{(3)}$ is defined within the foliation $\Sigma$. This provides a connection between the thermodynamics of a system expressed with the information current, and fluctuations from $\cmat{R}_{\Sigma}$. 

To evaluate $\cmat{R}_{\Sigma_f}(x_1,x_2)$, we first find an expression for the retarded Green's function. 
Consider again the defining equation for the retarded Green's function
\begin{equation}
   \left( \cmat{M}^{\mu} \partial_{\mu} + \cmat{V} \right) \cmat{G}_R(x,x') = \cmat{I} \,\delta^{(4)}(x-x') .
\end{equation}
Choosing a foliation $\Sigma$ with past-directed unit normal $n^{\mu}$, the equation of motion can be decomposed as
\begin{equation}
    \left( n_{\mu} \cmat{M}^{\mu} \frac{d}{d\tau} + \cmat{M}^{\mu} \partial_{\mu}^{\perp} + \cmat{V} \right) \cmat{G}_R(x,x') = \cmat{I} \,\delta^{(4)}(x-x') ,
\end{equation}
where %
\begin{equation}
 \frac{d}{d\tau} = -n^{\mu} \partial_{\mu}, \qquad \partial^{\perp}_{\mu} \equiv (g^{\nu}_{\mu} + n^{\nu} n_{\mu}) \partial_{\nu} \equiv \Delta_{(n)\mu}^{\nu} \partial_{\nu} 
\end{equation}
are the derivatives parallel and orthogonal to $n^{\mu}$, respectively. Then, 
\begin{equation}\label{eq:dtauGR}
    \left( \cmat{I} \frac{d}{d\tau} + \cmat{F} \right) \cmat{G}_R(x,x') = [(n_{\mu} \cmat{M}^{\mu})^{-1}]\, \delta^{(4)}(x-x') ,
\end{equation}
where
\begin{equation}\label{eq:FAB}
    \cmat{F} \equiv \left(n_{\mu} \cmat{M}^{\mu}\right)^{-1} \left( \cmat{M}^{\mu} \partial^{\perp}_{\mu} + \cmat{V} \right) .
\end{equation}
By formally exponentiating the operator $\cmat{F}$ and integrating from $\tau_0$ to $\tau$, 
one finds the solution to Eq.\ \eqref{eq:dtauGR}:
\begin{equation}\label{eq:GRexplicit}
    \cmat{G}_R(x,x') 
     = \Theta (\tau - \tau')\, \mathcal{T}_+ \left(e^{-\int^\tau_{\tau'}d\tau''\, \cmat{F}(\tau'' , x) }\right) \cdot \left(n_{\mu}(x')\, \cmat{M}^{\mu}\right)^{-1} \,\delta^{(3)}(x-x') . 
\end{equation}
To keep our approach foliation independent, i.e., independent of how one chooses to separate space and time, with $n^\mu = n^\mu(x)$ in general, we consider a $\cmat{F}$ which may depend on spacetime coordinates, and employ a $\tau$-ordering operator $\mathcal{T}_\pm$, with contributions from increasing (decreasing) values of $\tau$  being ordered from right to left for $\mathcal{T}_+$ ($\mathcal{T}_-$).
Substituting Eq.\ \eqref{eq:GRexplicit} into Eq.\ \eqref{eq:RSigma}, we find
\begin{equation} \label{eq:RSigma-int}
    \begin{split}
        \cmat{R}_{\Sigma_f}(x_1,x_2)
     = 2 \int_{\tau_0}^{\tau_f} d\tau' \, &
      \mathcal{T}_+ \left(e^{-\int^\tau_{\tau'}d\tau''\, \cmat{F}(\tau'' , x_1) }\right)
      \left(n_{\mu}(x_1)\, \cmat{M}^{\mu}\right)^{-1}
      \cmat{Q}  \\ 
     & \times \left(n_{\nu}(x_2)\, \cmat{M}^{\dagger \nu}\right)^{-1}  \mathcal{T}_- \left(e^{-\int^\tau_{\tau'}d\tau''\, \cmat{F}^\dagger(\tau'' , x_2) }\right)\,\delta^{(3)}(x_1 - x_2) .
    \end{split}
\end{equation}
Assuming that $d \cmat{R}_{\Sigma_f} / d\tau_f \rightarrow 0$ as $\tau_f \rightarrow \infty$, the derivative gives
\begin{equation}
    2 \left(n_{\mu}\, \cmat{M}^{\mu}\right)^{-1}  \cmat{Q}  \left(n_{\nu}\, \cmat{M}^{\dagger \nu}\right)^{-1} - \, \cmat{F} \,\cmat{R}_{\infty} - \, \cmat{R}_{\infty}  \cmat{F}^{\dagger} = 0 .
\end{equation}
Substituting Eq.\ \eqref{R=nE}, this becomes 
\begin{equation} \label{fdt_1}
    2 \left(n_{\mu}\, \cmat{M}^{\mu}\right)^{-1}  \cmat{Q} \left(n_{\nu}\, \cmat{M}^{\dagger \nu}\right)^{-1} = \cmat{F} ( n_{\mu} \cmat{E}^{\mu} )^{-1} + ( n_{\mu} \cmat{E}^{\mu} )^{-1} \cmat{F}^{\dagger} .
\end{equation}
This is a statement of the fluctuation-dissipation theorem in our approach. 

We note that the result Eq.\ \eqref{fdt_1} can also be obtained with no resource to path integrals, simply by matching the two point function to its equilibrium value at late times. The path integral employed above was used to create a covariant generalization of the approach used by Fox and Uhlenbeck \cite{doi:10.1063/1.1693183}.   

\subsection{Relativistic Fox-Uhlenbeck approach}
\label{FoxUhlenbeck}

Using this construction, we identify two main approaches that can be used to determine the noise correlators. To see the first such approach, consider the equation of motion in the form
\begin{equation}
    \left( n_{\mu} \cmat{M}^{\mu} \frac{d}{d\tau} + \cmat{M}^{\mu} \partial_{\mu}^{\perp} + \cmat{V} \right) \delta \cvec{\phi} = \cvec{\Xi} ,
\end{equation}
Instead of calculating the fluctuations for $\cvec{\Xi}$ from this, we can redefine
\begin{equation}
    \cvec{\Xi}_{(FU)} \equiv [(n_{\mu} \cmat{M}^{\mu})^{-1}]\, \cvec{\Xi} ,
\end{equation}
such that
\begin{equation}
    \left( \cmat{I} \frac{d}{d\tau} + \cmat{F} \right) \delta \cvec{\phi} = \cvec{\Xi}_{(FU)} .
\end{equation}
This has precisely the same general structure as the equation of motion used to study non-relativistic fluctuations in \cite{doi:10.1063/1.1693183}, hence we call this the relativistic Fox-Uhlenbeck approach. 

Using this decomposition of the equations of motion, the 
noise correlator becomes
\begin{equation}
    \langle  \cvec{\Xi}_{(FU)}(x) \, \cvec{\Xi}_{(FU)} (x') \rangle = 2\, \cmat{Q}^{(FU)}\, ,
\end{equation}
with
\begin{equation}
    \cmat{Q}^{(FU)} = \left(n_{\mu}\, \cmat{M}^{\mu}\right)^{-1}  \cmat{Q}  \left(n_{\nu}\, \cmat{M}^{\dagger \nu}\right)^{-1} . 
\end{equation}
The noise correlator is then given by 
\begin{equation} \label{fdt_FU}
    2 \cmat{Q}^{(FU)} = \cmat{F} \cdot ( n_{\mu} \cmat{E}^{\mu} )^{-1} + ( n_{\mu} \cmat{E}^{\mu} )^{-1} \cdot \cmat{F}^{\dagger} ,
\end{equation}
which is a relativistic generalization of the main result of the seminal paper by Fox and Uhlenbeck \cite{doi:10.1063/1.1693183}.

Because of the redefinition of the stochastic source, the noise correlator and the on-shell equations of motion become dependent on the foliation. Since the same foliation-dependent matrix, $(n_{\mu} \cmat{M}^{\mu})^{-1}$, has been shifted out of the equations of motion and into the noise, the correlators of $\delta \cvec{\phi}$ will thus be equivalent to an approach in which no such shift is made. However, in order to find  explicitly foliation-independent correlators $\langle \delta \cvec{\phi} \delta \cvec{\phi} \rangle$, we need a noise correlator $\cmat{Q}$ that is also independent of foliation.

\subsection{Entropy production approach}
\label{Entropy_Production_Approach}

We will now derive a fluctuation-dissipation relation that is foliation independent by using the information current to construct the equations of motion. At this point, it is useful to invoke considerations on causality, well-posedness, and stability. As shown in \cite{Gavassino:2022roi}, it turns out that the conditions for  Eq.\ \eqref{stochastic_EoM} to be causal and symmetric-hyperbolic are that $ \tilde{{E}}^{\mu} \equiv \frac{1}{2} \delta \cvec{\phi}^T \, \cmat{M}^{\mu}\, \delta \cvec{\phi} $ satisfies precisely the same conditions that the information current $\cmat{E}^\mu$ must obey for the system to be covariantly stable following the Gibbs stability criterion \cite{Gavassino:2022roi}. Because an information current with these properties is unique (see \cite{Gavassino:2021kjm}), it follows that ${E}^\mu = \tilde{{E}}^\mu$, up to a multiplicative constant, for any $\delta \cvec{\phi}$. One can set the multiplicative constant to $1$ by appropriately rescaling $\cmat{M}$, $\cmat{V}$, and $\cmat{Q}$. Thus, $\cmat{M}^\mu + (\cmat{M}^\mu)^T = 2\,\cmat{E}^\mu$. 

Consider again Eq.\ \eqref{fdt_1}. Using Eq.\ \eqref{eq:FAB} 
we can write
\begin{equation}
    \left(\cmat{M}^{\mu} \partial^{\perp}_{\mu} + \cmat{V} \right) \cdot \left(n_\mu \cmat{E}^\mu\right)^{-1} \left(n_\beta \cmat{M}^\beta \right)  +  
  \left(n_\alpha \cmat{M}^\alpha \right) \left(n_\mu \cmat{E}^\mu\right)^{-1}  \left( \cmat{M}^{\mu} \partial^{\perp}_{\mu} + \cmat{V} \right)^\dagger   = 2\, \cmat{Q} .
\end{equation}
Assuming
\begin{equation} \label{nM=nE}
    n_{\mu} \cmat{E}^{\mu} = n_{\mu} \cmat{M}^{\mu} = n_{\mu} \cmat{M}^{\dagger \mu} , 
\end{equation}
the noise correlator takes the form 
\begin{equation}
\begin{split}
    2 \cmat{Q} & = \left( \Delta_{(n)\nu}^{\mu} \cmat{M}^{\nu} \partial_{\mu} + \cmat{V} \right) + \left( - \Delta_{(n)\nu}^{\mu} \cmat{M}^{\nu} \partial_{\mu} + \cmat{V}^{\dagger} \right) \\
    & = \cmat{V} + \cmat{V}^{\dagger} + \Delta_{(n)\nu}^{\mu} (\cmat{M}^{\nu} - \cmat{M}^{\dagger \nu}) \partial_{\mu} .
\end{split}
\end{equation}
If $\cmat{M}$ is symmetric, which must be the case if Eq.\ \eqref{nM=nE} holds, the last term manifestly vanishes. The fluctuation-dissipation relation thus becomes
\begin{equation} \label{fdt_foliation_independent}
    2 \cmat{Q} = \cmat{V} + \cmat{V}^{\dagger} . 
\end{equation}
This version of the fluctuation-dissipation relation is fully relativistic and independent of the foliation. We note that this result might seem independent of the information current, but it really depends on $\cmat{E}^{\mu}$ indirectly through the condition that $\cmat{M}^{\mu} = \cmat{E}^{\mu}$. 
This condition demands that the equation of motion, Eq.\ \eqref{stochastic_EoM}, is rescaled by the appropriate constant. 
This has no effect on the solutions of Eq.\ \eqref{stochastic_EoM} but introduces the background thermal scale $T$ into the noise correlator $\cmat{Q}$.

We further point out that from Eqs.\ \eqref{eq:defE} and \eqref{stochastic_EoM}, neglecting the contribution from $\cvec{\Xi}$, the entropy production rate is 
\begin{equation}
    - \frac{1}{2}\partial_\mu E^\mu =\frac{1}{2} \delta\cvec{\phi}^T\left( \cmat{V} + \cmat{V}^T\right)  \delta\cvec{\phi} \geq 0 . 
\end{equation}
Defining the entropy-production matrix $\cmatgreek{\sigma}\equiv \frac{1}{2}\left(\cmat{V} + \cmat{V}^T\right)$ our main result in Eq.\ \eqref{fdt_foliation_independent} acquires the even simpler form
\begin{equation}
    2 \cmat{Q} = 2 \cmatgreek{\sigma} ,
\end{equation}
which directly links stochastic noise and entropy production.

The result in Eq.\ \eqref{fdt_foliation_independent} was derived by making the assumption that Eq.\ \eqref{nM=nE} holds. We now ask, is this a good assumption? The answer, as well as some physical insight, can be obtained from \cite{Gavassino:2023odx}. In that paper, it was shown that for a causal dynamical system of the form
\begin{equation} \label{EoM_asymmetric}
    \left( \cmat{M}^{\mu} \partial_{\mu} + \cmat{V} \right) \delta \cvec{\phi} = 0 ,
\end{equation}
there always exists some matrix $\cmat{N}$ such that
\begin{equation}
    \cmat{N} \left( \cmat{M}^{\mu} \partial_{\mu} + \cmat{V} \right) \delta \cvec{\phi} = \left( \cmat{E}^{\mu} \partial_{\mu} + \cmatgreek{\sigma} + \tilde{\cmat{V}}_{\mathrm{asym}} \right) \delta \cvec{\phi} = 0 ,
\end{equation}
\begin{equation}
    \left( \cmat{E}^{\mu} \partial_{\mu} + \cmatgreek{\sigma} + \tilde{\cmat{V}}_{\mathrm{asym}} \right) \delta \cvec{\phi} = \cmat{N} \cvec{\Xi} .
\end{equation}
Defining
\begin{equation}
    \Tilde{\cmat{V}} = \cmatgreek{\sigma} + \tilde{\cmat{V}}_{\mathrm{asym}} , \:\: \Tilde{\cvec{\Xi}} = \cmat{N} \cvec{\Xi} ,
\end{equation}
the equation of motion takes the form 
\begin{equation} \label{EoM_from_E}
    \left( \cmat{E}^{\mu} \partial_{\mu} + \Tilde{\cmat{V}} \right) \delta \cvec{\phi} = \Tilde{\cvec{\Xi}} .
\end{equation}
This is equivalent to the general form, except $\cmat{M}^{\mu} \rightarrow \cmat{E}^{\mu}$, which is symmetric. Using Eq.\ \eqref{fdt_foliation_independent}, it follows that
\begin{equation} \label{Q_information_current}
    2 \Tilde{\cmat{Q}} = \Tilde{\cmat{V}} + \Tilde{\cmat{V}}^{\dagger} = 2\,\cmatgreek{\sigma} ,
\end{equation}
where $\Tilde{\cmat{Q}}$ is the two-point correlation function of $\Tilde{\cvec{\Xi}}$. So, if the equation of motion is written in the form of Eq.\ \eqref{EoM_from_E}, the noise correlator is precisely given by the entropy production. This is a succinct (and very elegant) form of the fluctuation-dissipation theorem in which the fluctuations are represented by the noise correlator $\tilde{\cmat{Q}}$ and the dissipation is represented by the entropy production $\cmatgreek{\sigma}$. Such a construction is always possible if the system is covariantly stable and equilibrium is the state with the largest entropy. This approach has the benefit of being constructed solely from the information current and entropy production, which involve quantities with clear thermodynamic interpretation. 

This result is similar to that employed in the study of non-relativistic fluctuations using the Onsager coefficients \cite{Onsager:1931jfa, Onsager:1931kxm, landau_statistical_1980}. While it might appear that we could have just used this from the start, it is important to note that this result requires \eqref{nM=nE}, which is the case if the equations of motion are written in terms of the information current, as in Eq.\ \eqref{EoM_from_E}. For a general set of equations of motion, the noise expressions will be transformed by some matrix with respect to this result. It is therefore essential, in the study of the fluctuating relativistic systems we are considering, to use the information current to obtain the noise correlator of Eq.\ \eqref{Q_information_current}.

\subsection{Foliation independence of physical correlators}  

The above formalism represents a relativistic generalization of the approach defined in Ref.~\cite{doi:10.1063/1.1693183}. Our construction has the benefit of being explicitly causal, stable, and observer-independent. As we will see below, when working with an acausal theory, such as relativistic Navier-Stokes hydrodynamics, the properties we have assumed will not hold, and the formalism cannot be applied. For instance, due to the relativity of simultaneity, the retarded Green's function constructed in one frame of reference will not be truly retarded for another observer, breaking the foliation independence of the noise correlator. 

That being said, as long as the underlying dynamics are causal and thermodynamically stable, both of the above approaches (relativistic Fox-Uhlenbeck and entropy-production) can be employed, with the same results for the correlators of physical quantities. In each case, the equation of motion takes the form 
\begin{equation}
    \cmat{D}\, \delta \cvec{\phi} = \cmat{S} \,\cvec{\xi} ,
\end{equation}
where $\cmat{D}$ is some differential operator and $\cmat{S} \cvec{\xi}$ is some scheme dependent noise. These partial differential equations are fully equivalent, all that changes is how $\cmat{D}$ is decomposed and what variables are absorbed into the noise $\cvec{\xi}$, expressed by $\cmat{S}$. The correlator for the physical variables $\delta \cvec{\phi}$ is then obtained by taking
\begin{equation}
    \langle \delta \cvec{\phi} \,\delta \cvec{\phi}^T \rangle = \cmat{D}^{-1} \cmat{S} \left\langle \cvec{\xi}\, \cvec{\xi}^T \right\rangle \cmat{S}^{\dagger} \cmat{D}^{\dagger - 1} . 
\end{equation}
This will typically be done in momentum space. If we were to choose a different scheme, changing to some $\cmat{D}', \cmat{S}'$, the equations of motion remain equivalent as long as 
\begin{equation}
    \cmat{D}^{-1} \cmat{S} = \cmat{D}'^{-1} \cmat{S}' . 
\end{equation}
Thus, the correlator of $\delta \cvec{\phi}$ remains unchanged. This indicates that the two approaches for determining the correlation functions each lead to equivalent physics and, therefore, are all independent of the foliation, i.e., of the choice of $n^\mu(x)$,  as desired. That is, in the case of the relativistic Fox-Uhlenbeck approach, the  dependence of the noise correlator and that of the differential operator $\cmat{D}$ 
 on $n^\mu(x)$ will cancel out when taking averages involving physical variables.

\section{Fluctuations in a model of relativistic heat transport}
\label{Heat_equation_causal}

A simple example of a fluctuating system is a diffusive process involving only a single scalar degree of freedom.  We start with the diffusion equation written in a covariant manner,
 \begin{equation}
     u^{\mu} \partial_{\mu} \delta T = D \Delta^{\mu\nu} \partial_{\mu} \partial_{\nu} \delta T ,
 \end{equation}
 where $D$ is a diffusion constant, $T$ is the equilibrium temperature, and $\delta T$ represents temperature perturbations around equilibrium (the 4-velocity of the medium is taken to be constant). This is the archetypal example of a parabolic partial differential equation \cite{ChoquetBruhatGRBook}, and this feature does not change when writing the equation covariantly. To have sensible relativistic dynamics with a well-posed initial value problem, we would like to employ a hyperbolic generalization of this so that signals can be restricted to propagate in a causal manner \cite{ChoquetBruhatGRBook}. This is provided by the Cattaneo model
 \begin{equation} \label{Cattaneo_heat_diffusion}
     u^{\mu} \partial_{\mu} \delta T + \tau u^{\mu} u^{\nu} \partial_{\mu} \partial_{\nu} \delta T = D \Delta^{\mu\nu} \partial_{\mu} \partial_{\nu} \delta T ,
 \end{equation}
 where $\tau$ is a relaxation time \cite{1571698600732865152}.

To add stochastic fluctuations to this system using the framework developed in Sec.\ \ref{relativistic_fluctuations} we must write the equation of motion in a first order form. We thus introduce the transverse vector flux $q^{\mu}$ defined by
 \begin{equation}
     \delta q^{\mu} + \tau \, u^{\nu} \partial_{\nu} \delta q^{\mu} = -\kappa \Delta^{\mu\nu} \partial_{\nu} \delta T .
 \end{equation}
 This can be supplemented with 
 \begin{equation}
     a\, u^{\mu} \partial_{\mu} \delta T + \Delta^{\mu}_{\:\:\:\nu} \partial_{\mu} \delta q^{\nu} = 0 ,
 \end{equation}
 in which case we recover Eq.\ \eqref{Cattaneo_heat_diffusion} with $D = \kappa/a$, and $\kappa$ is the thermal conductivity. The information current for such a theory can be determined solely by the symmetries \cite{Gavassino:2022roi}, which gives 
 \begin{equation} \label{Heat_E}
     E^{\mu} = \left( \frac{a}{2 T^2} \delta T^2 + \frac{\beta_1}{2 T} \delta q^2 \right) u^{\mu} + \frac{1}{T^2} \delta T \delta q^{\mu} ,
 \end{equation}
 where $\beta_1\,\kappa T \equiv \tau$.
 One can immediately see the importance of the relaxation time $\tau$. 
 If we were to set $\tau = 0$ then there would be no term that goes with $\delta q^2$, preventing $n_\mu E^\mu$ from being cast as a sum of squares for arbitrary timelike $n^\mu$, which leads to the issues discussed in Sec.\ \ref{navier_stokes_fluctuations}. This is precisely the sort of issue that one would find if they attempted to fluctuate relativistic Navier-Stokes in the Eckart frame with nonzero heat flux. Projecting Eq.\ \eqref{Heat_E} along an arbitrary $n_{\mu}$, we find that 
 \begin{equation}
     E = n_{\mu} E^{\mu} = \left( \frac{a}{2 T^2} \delta T^2 + \frac{\beta_1}{2 T} \delta q^2 \right) n_{\mu} u^{\mu} + \frac{1}{T^2} \delta T \,n_{\mu} \delta q^{\mu} ,
 \end{equation}
 which can be written in matrix form as %
 \begin{equation}
     \cmat{E} = \frac{1}{T^2} \begin{pmatrix}
         a\, n_{\lambda} u^{\lambda} & n^{\lambda}\Delta_{\nu\lambda} \\
         n^{\lambda}\Delta^\mu_\lambda & \beta_1 T\, n_{\lambda} u^{\lambda} \Delta^{\mu}_{\:\:\nu}
     \end{pmatrix} .
 \end{equation}
 Note that this is written in a block form, hence the free $\mu$ indices in the off-diagonal blocks.

 \subsection{Fox-Uhlenbeck approach}
 \label{Cattaneo_FU}

 We now choose to take $n^{\mu} = -u^{\mu}$ and work in the local-rest-frame $u^{\mu} = (1,0,0,0)$, such that 
 \begin{equation}
     \cmat{E} = \frac{1}{T^2} \begin{pmatrix}
         a & 0 \\
         0 & \beta_1 T \, \delta^i_j
     \end{pmatrix} ,
 \end{equation}
 and
 \begin{equation}
     \frac{d}{d\tau} = -n^{\mu} \partial_{\mu} = u^{\mu} \partial_{\mu} = \partial_t .
 \end{equation}
 Note that we have reduced $\cmat{E}$ from a five-by-five matrix to a four-by-four matrix because $\delta q^{\mu}$ is transverse to $u^{\mu}$, and so in the rest frame the only nonzero components are $\delta q^i$. The equations of motion now take the form
 \begin{equation}
     \partial_t \delta T + \frac{1}{a} \partial_i \delta q^i = \xi_{T}
 \end{equation}
 \begin{equation}\label{EoMq}
     \partial_t \delta q^i + \frac{1}{\tau} \delta q^i + \frac{1}{\beta_1 T} \partial^i \delta T = \xi_{q}^i ,
 \end{equation}
 where we have inserted stochastic terms $\xi_{T}, \xi_q^i$. We can Fourier transform the spatial derivatives, in which case this is in the form of Eq.\ \eqref{stochastic_EoM} with
 \begin{equation}
     \cmat{F} = \begin{pmatrix}
         0 & \frac{i k_j}{a} \\
         \frac{i k^i}{\beta_1 T} & \frac{1}{\tau} \delta^i_j 
     \end{pmatrix} .
 \end{equation}
 This gives everything needed to determine the noise correlators with Eq.\ \eqref{fdt_FU}, in which case we find that the only nonzero noise correlator is 
 \begin{equation} \label{Cattaneo_noise_correlator}
     \langle \xi_q^{\mu}(x) \xi_q^{\nu}(x') \rangle = \frac{2}{\kappa \beta_1^2} \Delta^{\mu\nu} \delta^{(4)}(x-x') .
 \end{equation}
 Using this and the equation of motion, we can determine the momentum-space correlators of the thermodynamic variables $\delta T, \delta q^i$ to be 
 \begin{equation}
     \langle \delta T(k^{\mu}) \delta T(-k^{\mu}) \rangle = \frac{2 T^3 \tau \beta_1 k^2}{k^4 \tau^2 + a T \omega^2 \beta_1 \left( -2k^2 \tau^2 + a \beta_1 T (1 + \omega^2 \tau^2) \right)}
 \end{equation}
 \begin{equation}
     \langle \delta T(k^{\mu}) \delta q^i(-k^{\mu}) \rangle = \frac{2a T^3 \tau \beta_1 \omega k^i}{k^4 \tau^2 + a T \omega^2 \beta_1 \left( -2k^2 \tau^2 + a \beta_1 T (1 + \omega^2 \tau^2) \right)}
 \end{equation}
 \begin{equation}
 \begin{split}
     \langle \delta q^i(k^{\mu}) \delta q^j(-k^{\mu}) \rangle = & \frac{2a^2 T^3 \tau \beta_1 \omega^2}{k^4 \tau^2 + a T \omega^2 \beta_1 \left( -2k^2 \tau^2 + a \beta_1 T (1 + \omega^2 \tau^2) \right)} \frac{k^ik^j}{k^2} + \\
     & + \frac{2 \tau T}{\beta_1 + \tau^2 \beta_1 \omega^2} \Delta_{(k)}^{ij} ,
\end{split}
 \end{equation}
 where $\Delta_{(k)}^{ij}$ is the projector orthogonal to $k^i$ in the spatial direction and $\omega$ is the Fourier conjugate to $t$. Using this new theory of fluctuations, we were able to systematically determine the stochastic correlators of both noise and hydrodynamic variables in Cattaneo's theory of heat transport. 

 Now, consider the case of shear modes in relativistic Navier-Stokes theory in the Landau-Lifshitz frame. This corresponds to setting $\tau = 0$ and taking $\delta T \rightarrow \delta \phi$, where $\delta \phi$ represents the shear component of the fluid four-velocity. The information current is then given by
 \begin{equation}
     \cmat{E} = \frac{1}{\phi^2} \begin{pmatrix}
         a n_{\lambda} u^{\lambda} & n_{\nu} \\
         n^{\mu} & 0
     \end{pmatrix} ,
 \end{equation}
 This matrix has zero determinant and therefore is not invertible. However, Eq.\ \eqref{fdt_FU} for the noise correlator involves the inverse of $\cmat{E}$. This implies that we cannot determine the fluctuations for relativistic Navier-Stokes using this framework. This might seem like a drawback, but it is actually a key feature of using the information current to determine fluctuations. When the equilibrium state is not the state with maximal entropy, and the system is not covariantly stable, this construction shows that the corresponding thermal fluctuations do not display the same properties for all inertial reference frames.

 \subsection{Entropy production approach}

For comparison, we now  use the approach from Sec.\ \ref{Entropy_Production_Approach} to determine noise correlators. Writing the equations of motion for the stochastic Cattaneo model in the form 
 \begin{equation}
     \left( \cmat{M}^{\mu} \partial_{\mu} + \cmat{V} \right) \delta \cvec{\phi} = \cvec{\Xi} , 
 \end{equation}
 with $\delta \cvec\phi = \{ \delta T, \delta q^{\mu} \}$, we find that
 \begin{equation}
     \cmat{M}^{\mu} = \begin{pmatrix}
         au^{\mu} & \Delta^{\mu}_{\lambda} \\
         \Delta^{\mu\nu} & \beta_1 T u^{\mu} \Delta^{\nu}_{\lambda} 
     \end{pmatrix} ,
 \end{equation}
 \begin{equation}
     \cmat{V} = \frac{1}{\kappa} \begin{pmatrix}
         0 & 0 \\
         0 & \Delta^{\mu}_{\nu} 
     \end{pmatrix} .
 \end{equation}
 It is simple to rescale the equations of motion such that $\cmat{M}^{\mu} = \cmat{E}^{\mu}$, so we recall
 \begin{equation}
     \cmat{E}^{\mu} = \frac{1}{T^2} \begin{pmatrix}
         au^{\mu} & \Delta^{\mu}_{\lambda} \\
         \Delta^{\mu\nu} & \beta_1 T u^{\mu} \Delta^{\nu}_{\lambda} 
     \end{pmatrix} .
 \end{equation}
 We can then rescale $\cmat{M}^{\mu}$ by $1 / T^2$, in which case
 \begin{equation}
      \cmat{M}^{\mu} = \frac{1}{T^2} \begin{pmatrix}
         au^{\mu} & \Delta^{\mu}_{\lambda} \\
         \Delta^{\mu\nu} & \beta_1 T u^{\mu} \Delta^{\nu}_{\lambda} 
     \end{pmatrix} ,
 \end{equation}
 \begin{equation}
     \cmat{V} = \frac{1}{\kappa T^2} \begin{pmatrix}
         0 & 0 \\
         0 & \Delta^{\mu}_{\nu}
     \end{pmatrix} .
 \end{equation}
 The noise correlator is then given by 
 \begin{equation}
     2 \cmat{Q} = \left( \cmat{V} + \cmat{V}^{\dagger} \right) = \frac{2}{\kappa T^2} \begin{pmatrix}
         0 & 0 \\
         0 & \Delta^{\mu}_{\nu}
     \end{pmatrix} .
 \end{equation}
 The noise in this form is manifestly independent of $n^{\mu}$ (it never even appeared in the calculation). Note also that this implies that the entropy production of the Cattaneo model is given by 
 \begin{equation}
     \sigma = \frac{1}{\kappa T^2} \Delta_{\mu\nu} \delta q^{\mu} \delta q^{\nu} .
 \end{equation}
 We can thus see that $\delta T$ is a conserved scalar, while $\delta q^{\mu}$ is the dissipative term. 

 To compare this to the earlier derivation using the foliation-dependent approach, we use that 
 \begin{equation}
     \cmat{Q}^{(FU)} = (n_{\mu} \cmat{M}^{\mu})^{-1} \cmat{Q} (n_{\mu} \cmat{M}^{\dagger \mu})^{-1} . 
 \end{equation}
 Using $n^{\mu} = -u^{\mu} = (-1,0,0,0)$, we find that 
 \begin{equation}
     \cmat{Q}^{(FU)} = \frac{2}{\kappa \beta_1^2} \begin{pmatrix}
         0 & 0 \\
         0 & \Delta^{\mu}_{\nu}
     \end{pmatrix} ,
 \end{equation}
which is precisely the result obtained in Eq.\ \eqref{Cattaneo_noise_correlator}. The correlator for $\delta \cvec{\phi}$ will take the same form as was obtained in the previous section.

\section{Fluctuations in conformal Israel-Stewart theory in a general hydrodynamic frame}
\label{gIS_fluctuations}

Now that we have considered a simple example, we will compute the fluctuations for Israel-Stewart theory in a general hydrodynamic frame (gIS) \cite{Noronha:2021syv}. The conformal limit is considered for simplicity of presentation and analysis, but the techniques presented here will work for the more general case as well. Generalized Israel-Stewart theory in the conformal limit describes the hydrodynamic evolution of a physical system defined by 
\begin{equation} \label{gIS_Tmunu}
    T^{\mu\nu} = (\epsilon + \mathcal{A}) \left( u^{\mu} u^{\nu} + \frac{1}{3} \Delta^{\mu\nu} \right) + u^{\mu} Q^{\nu} + u^{\nu} Q^{\mu} + \pi^{\mu\nu} ,
\end{equation}
where $\mathcal{A}$ is the out-of-equilibrium correction to energy density, $Q^{\mu}$ is the energy diffusion, and $\pi^{\mu\nu}$ is the shear-stress tensor. The conservation equations are then given by 
\begin{equation} \label{gIS_eom1}
    \mathcal{D} (\epsilon + \mathcal{A}) + \pi_{\mu\nu} \sigma^{\mu\nu} + \mathcal{D}_{\mu} Q^{\mu} = 0
\end{equation}
\begin{equation} \label{gIS_eom2}
    \Delta^{\lambda\nu} \left( \frac{1}{3} \mathcal{D}_{\lambda} (\epsilon + \mathcal{A}) \right) + Q^{\mu} \mathcal{D}_{\mu} u^{\nu} + \mathcal{D} Q^{\nu} = 0 ,
\end{equation}
where $\sigma^{\mu\nu} = \mathcal{D}^{\mu} u^{\nu} + \mathcal{D}^{\nu} u^{\mu}$, $\mathcal{D}_{\mu}$ is the Weyl-covariant derivative defined in \cite{Loganayagam:2008is}, and $\mathcal{D} = u^{\mu} \mathcal{D}_{\mu}$. Here, $\mathcal{A},Q^{\mu}, \pi^{\mu\nu}$ are treated as new hydrodynamic degrees-of-freedom that obey the relaxation equations 
\begin{equation} \label{gIS_relax_pi}
    \tau_{\pi} \left[ \mathcal{D} \pi^{\mu\nu} + \frac{1}{2} \pi^{\mu\nu} \mathcal{D} \ln \left( \frac{\tau_{\pi}}{\eta T} \right) \right] + \pi^{\mu\nu} = -2\eta \sigma^{\mu\nu} 
\end{equation}
\begin{equation} \label{gIS_relax_Q}
    \tau_Q \left[ \mathcal{D} Q^{\mu} + \frac{1}{2} Q^{\mu} \mathcal{D} \ln \left( \frac{\tau_Q}{\epsilon \tau_{\psi} T} \right) \right] + Q^{\mu} = -\tau_{\psi} \Delta^{\mu\lambda} \mathcal{D}_{\lambda} \epsilon
\end{equation}
\begin{equation} \label{gIS_relax_A}
    \tau_A \left[ \mathcal{D} \mathcal{A} + \frac{1}{2} \mathcal{A} \mathcal{D} \ln \left( \frac{\tau_A}{\epsilon \tau_{\phi} T} \right) \right] + \mathcal{A} = -\tau_{\phi} \mathcal{D} \epsilon,
\end{equation}
which were obtained by imposing that the entropy production is non-negative \cite{Noronha:2021syv}. Above, $\eta$ is the shear viscosity and $\{\tau_Q, \tau_A, \tau_{\psi}, \tau_{\phi},\tau_\pi\}$ have units of relaxation time ($\sim 1/T$). 
We note that in the Landau-Lifshitz frame, only the first of these relaxation equations is present, hence we think of this frame as corresponding to $\tau_Q = \tau_A = \tau_{\psi} = \tau_{\phi} = 0$. In general, however, the additional two relaxation equations are allowed by the symmetries of the energy-momentum tensor and the second law of thermodynamics. The final necessary expression to determine the fluctuations of gIS is the entropy current, which is given by
\begin{equation} \label{gIS_entropy}
    s^{\mu} = \left( s + \frac{\mathcal{A}}{T} \right) u^{\mu} + \frac{Q^{\mu}}{T} - \frac{1}{2T} \left( \frac{\tau_{\pi}}{\eta} \pi_{\alpha\beta} \pi^{\alpha\beta} + \frac{\tau_Q}{4\epsilon \tau_{\psi}} Q_{\lambda} Q^{\lambda} + \frac{\tau_A}{4\epsilon \tau_{\phi}} \mathcal{A}^2 \right) u^{\mu}.
\end{equation}

\subsection{Entropy production approach}
\label{gIS_foliation_independent}

The simplest way to introduce stochastic fluctuations in gIS is to use the information current approach. The information current for gIS is found to be
\begin{equation}
\begin{split}
    E^{\mu} & = \frac{2 \epsilon u^{\mu}}{T^3} \delta T^2 + \frac{2 \epsilon u^{\mu}}{3T} \delta u^{\nu} \delta u_{\nu} + \frac{c_V \delta T \delta u^{\mu}}{3T} + \frac{\delta \mathcal{A} \delta u^{\mu}}{3T} + \frac{u^{\mu}}{T} \delta u^{\nu} \delta Q_{\nu} + \\
    & + \frac{\delta u_{\nu} \delta \pi^{\mu\nu}}{T} + \frac{u^{\mu}}{T^2} \delta \mathcal{A} \delta T + \frac{\delta T \delta Q^{\mu}}{T^2} + \frac{u^{\mu}}{2T} \left[ \frac{\tau_A}{4\epsilon \tau_{\phi}} \delta \mathcal{A}^2 + \frac{\tau_Q}{4 \epsilon \tau_{\psi}} \delta Q^2 + \frac{\tau_{\pi}}{\eta} \delta \pi^2 \right] .
\end{split}
\end{equation}

Using the approach described in Sec.\ \ref{Entropy_Production_Approach}, the equation of motion can then be rewritten in the form 
\begin{equation}
    \left( \cmat{E}^{\mu} \partial_{\mu} + \cmatgreek{\sigma} + \tilde{\cmat{V}}_{\mathrm{asym}} \right) \delta \cvec{\phi} = \cvec{\Xi} ,
\end{equation}
where 
\begin{equation}
    \sigma = - \mathcal{D}_{\mu} E^{\mu} ,
\end{equation}
and the space of thermodynamic variables is taken to be $\delta \cvec{\phi} = \{ 
\delta \epsilon, \delta \mathcal{A}, (\epsilon + P) \delta u^{\mu}, \delta Q^{\mu}, \delta \pi^{\mu\nu} \}$, and $\tilde{\cmat{V}}_{\mathrm{asym}}$ is determined from the equations of motion to be $\tilde{\cmat{V}}_{\mathrm{asym}} = 0$. The noise correlator is then given by 
\begin{equation}
    2 \cmat{Q} = \cmatgreek{\sigma} + \cmatgreek{\sigma}^{\dagger} = 2 \cmatgreek{\sigma} .
\end{equation}
To determine the foliation-independent noise all that is needed is to determine $\cmatgreek{\sigma}$.

To calculate $\cmatgreek{\sigma}$, recall that the information current is given by 
\begin{equation}
    E^{\mu} = -\delta s^{\mu} - \beta_{\nu} \delta T^{\mu\nu} .
\end{equation}
The Killing vector $\beta_{\nu}$ is calculated in the background equilibrium fluid, so its derivative is zero, while $\mathcal{D}_{\mu} \delta T^{\mu\nu} = 0$ from conservation of energy and momentum. It follows that
\begin{equation}
    \sigma = - \mathcal{D}_{\mu} E^{\mu} = \mathcal{D}_{\mu} \delta s^{\mu} ,
\end{equation}
which is simply the entropy production of gIS, as expected. This was calculated in \cite{Noronha:2021syv}, and the result is
\begin{equation}
    \sigma = \mathcal{D}_{\mu} \delta s^{\mu} = \frac{1}{2T} \left( \frac{\delta \mathcal{A}^2}{4\epsilon \tau_{\phi}} + \frac{\delta Q_{\mu} \delta Q^{\mu}}{4\epsilon \tau_{\psi}} + \frac{\delta \pi_{\mu\nu} \delta \pi^{\mu\nu}}{\eta} \right) . 
\end{equation}
The noise correlator is then given by 
\begin{equation}
    2 \cmat{Q} = \begin{pmatrix}
        0 & 0 & 0 & 0 & 0 \\
        0 & \frac{1}{4\epsilon T \tau_{\phi}} & 0 & 0 & 0 \\
        0 & 0 & 0 & 0 & 0 \\
        0 & 0 & 0 & \frac{1}{4 \epsilon T \tau_{\psi}} \Delta_{\nu}^{\mu} & 0 \\
        0 & 0 & 0 & 0 & \frac{1}{\eta T} \Delta_{\alpha\beta}^{\mu\nu}
    \end{pmatrix} . 
\end{equation}
This can now be used to determine the symmetrized correlator of the energy-momentum tensor. Note that the correlators for the noise components $\Xi_{\epsilon}$ and $\Xi_{u}^{\mu}$ are both identically zero. This result was not imposed -- it comes out from the formalism.  In the end, this is to be expected since those were the stochastic terms present in the conservation law $\partial_{\mu} T^{\mu\nu} = 0$.

\subsection{Correlator of energy-momentum disturbances}
\label{Tmunu_corr}

The equation of motion for gIS can be written in the form
\begin{equation}
    \cmat{D} \delta \cvec{\phi} = \cvec{\Xi} ,
\end{equation}
where
\begin{equation}
    \cmat{D}(\omega,k) = -i\omega \cmat{E}^0 + i k_i \cmat{E}^i + \cmatgreek{\sigma} 
\end{equation}
in momentum space. This can be inverted to find that 
\begin{equation}\label{fluct_from_noise}
    \delta \cvec{\phi} = \cmat{D}^{-1} \cvec{\Xi} .
\end{equation}
Squaring Eq.\ \eqref{fluct_from_noise} and taking the expectation value, we obtain 
\begin{equation} \label{corr_hydro_vars}
    \langle \delta \cvec{\phi}(k^{\mu}) \delta \cvec{\phi}^T(-k^{\mu}) \rangle = \cmat{D}^{-1}(\omega,k) \cmat{Q} \cmat{D}(\omega,k)^{\dagger -1}.
\end{equation}
In principle, this expression is all that is necessary to determine the correlator of energy-momentum disturbances, but the symmetries can be exploited to simplify the calculations. 

Working in momentum space, we can determine the correlator of energy-momentum disturbances by following the decomposition used in \cite{Kovtun:2005ev}. We thus define the projector orthogonal to the four-momentum as
\begin{equation}
    \Delta_{(k)}^{\mu\nu} = g^{\mu\nu} - \frac{k^{\mu} k^{\nu}}{k^{\lambda} k_{\lambda}} .
\end{equation}
This projector can be decomposed into transverse and longitudinal parts as $\Delta^{\mu\nu}_{(k)} = P_{\perp}^{\mu\nu} + P_L^{\mu\nu}$, which in the local rest frame satisfy
\begin{equation}
    P_{\perp}^{0\mu} = 0, \:\: P_{\perp}^{ij} = \delta^{ij} - \frac{k^ik^j}{k^2} ,
\end{equation}
and
\begin{equation}
    P_L^{00} = \frac{k^2}{\omega^2 - k^2}, \:\: P_L^{0i} = \frac{\omega k^i}{\omega^2 - k^2}, \:\: P_L^{ij} = \frac{\omega^2}{k^2} \frac{k^i k^j}{\omega^2 - k^2} .
\end{equation}
In terms of these projectors, the energy-momentum correlator can be written as
\begin{equation}\label{eq:genTmunu-fluct}
    \langle \delta T^{\mu\nu}(k^{\lambda}) \delta T^{\alpha\beta}(-k^{\lambda}) \rangle = G_1(\omega,k^2) S^{\mu\nu,\alpha\beta} + G_2(\omega,k^2) Q^{\mu\nu,\alpha\beta} + G_3(\omega,k^2) L^{\mu\nu,\alpha\beta} .
\end{equation}
The tensor structure of the correlator is defined by the rank-4 orthogonal projectors
\begin{equation}
    S^{\mu\nu,\alpha\beta} = \frac{1}{2} \left( P_{\perp}^{\mu\alpha} P_L^{\nu\beta} + P_{L}^{\mu\alpha} P_{\perp}^{\nu\beta} + P_{\perp}^{\mu\beta} P_L^{\nu\alpha} + P_{L}^{\mu\beta} P_{\perp}^{\nu\alpha} \right) ,
\end{equation}
\begin{equation}
    Q^{\mu\nu,\alpha\beta} = \frac{1}{3} \left[ P_{L}^{\mu\nu} P_L^{\alpha\beta} + \frac{1}{2} P_{\perp}^{\mu\nu} P_{\perp}^{\alpha\beta} - \left( P_{\perp}^{\mu\nu} P_L^{\alpha\beta} + P_L^{\mu\nu} P_{\perp}^{\alpha\beta} \right) \right] ,
\end{equation}
\begin{equation}
    L^{\mu\nu,\alpha\beta} = \Delta^{\mu\nu\alpha\beta} - S^{\mu\nu,\alpha\beta} - Q^{\mu\nu,\alpha\beta} .
\end{equation}
Since there are only three structures present, we can pick three components of each of these tensors and use them to extract the symmetrized correlator of the energy-momentum tensor. 

Working in the local rest frame and taking the momentum to be in the $x$-direction, that is $k^1 = k$, we find that 
\begin{equation}
    S^{00,00} = 0, \:\: S^{03,03} = \frac{1}{2} \frac{k^2}{\omega^2 - k^2}, \:\: S^{12,12} = 0
\end{equation}
\begin{equation}
    Q^{00,00} = \frac{2}{3} \frac{k^4}{(\omega^2 - k^2)^2} , \:\: Q^{03,03} = 0, \:\: Q^{12,12} = 0
\end{equation}
\begin{equation}
    L^{00,00} = 0, \:\: L^{03,03} = 0, \:\: L^{12,12} = \frac{1}{2} \frac{\omega^2}{\omega^2 - k^2} .
\end{equation}
We can thus simply take these three components of $\langle T^{\mu\nu}(k^{\lambda}) T^{\alpha\beta}(-k^{\lambda}) \rangle$ to extract the coefficients $G_i(\omega,k^2)$. We therefore compute 
\begin{equation}
    \langle \delta T^{00}(k^{\mu}) \delta T^{00}(-k^{\mu}) \rangle = \langle \delta \epsilon(k^{\mu}) \delta \epsilon(-k^{\mu}) \rangle + \langle \delta \mathcal{A}(k^{\mu}) \delta \mathcal{A}(-k^{\mu}) \rangle + 2 \langle \delta \epsilon(k^{\mu}) \delta \mathcal{A}(-k^{\mu}) \rangle ,
\end{equation}
\begin{equation}
\begin{split}
    \langle \delta T^{03}(k^{\mu}) \delta T^{03}(-k^{\mu}) \rangle = & \langle (\epsilon + P) \delta u^3(k^{\mu}) (\epsilon + P) \delta u^3(-k^{\mu}) \rangle + \langle \delta Q^3(k^{\mu}) \delta Q^3(-k^{\mu}) \rangle + \\
    & + 2 \langle (\epsilon + P) \delta u^3(k^{\mu}) \delta Q^3(-k^{\mu}) \rangle ,
\end{split}
\end{equation}
\begin{equation}
    \langle \delta T^{12}(k^{\mu}) \delta T^{12}(-k^{\mu}) \rangle = \langle \delta \pi^{12}(k^{\mu}) \delta \pi^{12}(-k^{\mu}) \rangle .
\end{equation}
The energy-momentum correlator is then defined by 
\begin{equation}
    G_1(\omega,k^2) = \frac{2(\omega^2 - k^2)}{k^2} \bigg[ \langle (\epsilon + P) \delta u^3 (\epsilon + P) \delta u^3 \rangle + \langle \delta Q^3 \delta Q^3 \rangle + 2 \langle (\epsilon + P) \delta u^3 \delta Q^3 \rangle \bigg] ,
\end{equation}
\begin{equation}
    G_2(\omega,k^2) = \frac{3}{2} \frac{(\omega^2 - k^2)^2}{k^4} \bigg[ \langle \delta \epsilon \delta \epsilon \rangle + \langle \delta \mathcal{A} \delta \mathcal{A} \rangle + 2 \langle \delta \epsilon \delta \mathcal{A} \rangle  \bigg] ,
\end{equation}
\begin{equation}
    G_3(\omega,k^2) = \frac{2(\omega^2 - k^2)}{\omega^2} \langle \delta \pi^{12} \delta \pi^{12} \rangle .
\end{equation}
Using Eq.\ \eqref{corr_hydro_vars}, each of the correlators of hydrodynamic variables present in these expressions can be computed. Doing so and defining $\tau_{\eta} = \eta / (\epsilon + P)$, we obtain
\begin{equation}
    \begin{split}
        G_1(\omega,k^2) & = - \Big\{ 16 T \epsilon (\omega^2 - k^2) \tau_{\eta} \left[ 1 + \omega^2 \tau_Q^2 - 6\omega^2 \tau_Q \tau_{\psi} + 3 k^2 \tau_{\eta} \tau_{\psi} + 9 \omega^2 \tau_{\psi}^2 \right] \Big\} \times \\
        & \times \Big\{ 3 \left[ (1 + \omega^2 \tau_Q^2) (\omega^2 + (\omega^2 \tau_{\pi} - k^2 \tau_{\eta})^2) \right] + 6\omega^2 \left[ k^2 \tau_{\eta} - \omega^2 \tau_Q (1 + \omega^2 \tau_{\pi}^2 - k^2 \tau_{\eta} \tau_{\pi}) \right] \tau_{\psi} + \\
        & + 9\omega^4 (1 + \omega^2 \tau_{\pi}^2) \tau_{\psi}^2 \Big\}^{-1} ,
    \end{split}
\end{equation}
\begin{equation}
    G_2(\omega, k^2) = \frac{f(\omega,k^2)}{|D(\omega, k^2)|^2 } ,
\end{equation}
where 
\begin{equation}
    \begin{split}
        f(\omega, k^2) & = 48 T \epsilon  \tau _{\eta } (\omega^2 - k^2)^2 \Big\{ 2 \tau _{\psi } \big[ k^2 \big( 2 \tau _{\eta } [ \omega ^2 ( \tau _A-\tau _{\phi } ){}^2+1 ] + \tau _{\phi } \big)-\omega ^2 \tau _Q \big( (\tau _A-\tau _{\phi }) \times \\
        & \times [ 3 \omega ^2 \tau _A+ (k^2-3 \omega ^2 ) \tau _{\phi } ] +3\big) \big] + \tau _{\psi }^2 [ 6 \omega ^2 \tau _A ( k^2-3 \omega ^2 ) \tau _{\phi }+9 \omega ^4 \tau _A^2+4 k^4 \tau _{\eta } \tau _{\phi } + \\
        & + ( k^2-3 \omega ^2 )^2 \tau _{\phi }^2+9 \omega ^2 ] + ( \omega ^2 \tau _Q^2+1 ) [ \omega ^2 ( \tau _A-\tau _{\phi } ){}^2+1 ] \Big\} ,
    \end{split}
\end{equation}
\begin{equation}
    \begin{split}
        D(\omega, k^2) & = \omega  \tau _A \big( ( \omega  \tau _Q-i ) [ -4 k^2 \omega  \tau _{\eta }-( k^2-3 \omega ^2 ) ( \tau _{\pi } \omega -i ) ]+ \tau _{\psi } [ 4 k^4 \tau _{\eta }+3 \omega  ( k^2-3 \omega ^2 ) \times \\
        & \times ( \tau _{\pi } \omega -i ) ] \big) + ( 1+i \omega  \tau _Q ) ( \omega  \tau _{\phi }+i ) ( -4 i k^2 \omega  \tau _{\eta }-k^2+3 \omega ^2 ) + \tau _{\pi } \omega  ( k^2-3 \omega ^2 ) \times \\
        & \times \big(\tau _{\psi } [ ( k^2-3 \omega ^2 ) \tau _{\phi }-3 i \omega ] + ( \omega  \tau _Q - i ) ( \omega  \tau _{\phi } + i ) \big) - i \tau _{\psi } [ 4 k^4 \tau _{\eta } + ( k^2 - 3 \omega ^2 )^2 \tau _{\phi } + \\
        & - 3 i k^2 \omega +9 i \omega ^3 ] ,
    \end{split}
\end{equation}
and
\begin{equation}
    \begin{split}
        G_3(\omega,k^2) & = \left\{16 T (\omega ^2 - k^2) \epsilon  \tau _{\eta } \left(3 k^2 \tau _{\eta } \tau _{\psi }-6 \omega ^2 \tau _Q \tau _{\psi }+\omega ^2 \tau _Q^2+9 \omega ^2 \tau _{\psi }^2+1\right) \right\} \times \\
        & \times \Big\{ 3 \Big[6 \omega ^2 \tau _{\psi } \left(k^2 \tau _{\eta }+\omega ^2 \tau _Q \left(k^2 \tau _{\pi } \tau _{\eta }-\tau _{\pi }^2 \omega ^2-1\right)\right)+\left(\omega ^2 \tau _Q^2+1\right) \left(\left(\tau _{\pi } \omega ^2-k^2 \tau _{\eta }\right){}^2+\omega ^2\right) + \\
        & + 9 \omega ^4 \tau _{\psi }^2 \left(\tau _{\pi }^2 \omega ^2+1\right)\Big] \Big\}^{-1} .
    \end{split}
\end{equation}
Note that as we take the limit of $k^2 \rightarrow 0$, each of these coefficients become
\begin{equation}
    G_1(\omega,0) = G_2(\omega,0) = G_3(\omega,0) = \frac{16 \epsilon T \tau_{\eta}}{3 (1 + \omega^2 \tau_{\pi}^2)} .
\end{equation}
This equality is expected from rotational invariance, making this an important check of the validity of this correlator.

{Equation \eqref{eq:genTmunu-fluct} presents the most general tensorial structure for the symmetrized correlator of the energy-momentum tensor allowed by symmetries. In the gIS theory, we find that the coefficients for each of these structures are independent and non-vanishing, so all of the allowed structures are present in the correlation function. This should be contrasted with the IS theory in the Landau-Lifshitz frame, in which these structures combine into a single term $\propto \Delta^{\mu\nu\alpha\beta}$ and, thus, there is only one independent structure in the two-point function. This difference stems from the fact that in a general frame, one has one off-equilibrium current for each component of $T^{\mu\nu}$. Because each off-equilibrium current introduces a stochastic noise source, this leads to the most general two-point function allowed by symmetries. In contrast, the IS theory in the Landau-Lifshitz frame has one degree of freedom for each component of $T^{\mu\nu}$ and the only off-equilibrium currents are the components of $\pi^{\mu\nu}$, leading to restrictions in how the system is allowed to fluctuate.}

{The observation that taking the Landau-Lifshitz frame restricts fluctuations of the energy-momentum tensor is not a mere technicality. This frame is an interesting choice in the IS theory precisely because it yields the most general decomposition for the one-point function of the energy-momentum tensor. However, imposing this choice \emph{off-shell} is stronger than imposing it on the one-point function, thus leading to a two-point function that is not fully general. } 

{This point can be illustrated with a simpler, yet analogous, situation. Suppose we have a spinning particle, and we choose the coordinate axis $\hat z$ to lie along its spin, with no loss of generality. Consider now that this system fluctuates: in principle, each of the three components of the angular momentum can fluctuate independently, so taking the spin of the particle to lie along the same axis \emph{at the level of fluctuations} will artificially remove one-third of the allowed fluctuations.}

\section{Conclusions}
\label{conclusions}

In this paper, we have constructed a new, general procedure for the inclusion of fluctuations in viscous relativistic hydrodynamics. 
Unlike other theories of fluctuating hydrodynamics available in the literature, our proposal incorporates recent developments on the stability and causality of relativistic fluids \cite{Gavassino:2021owo, Gavassino:2021kjm}.
The present theory of fluctuating hydrodynamics is realized in an observer-independent manner, in accordance with the principle 
of relativity, by foliating spacetime in a set of arbitrary spacelike hypersurfaces $\Sigma$. 
As we have shown, our approach avoids inconsistencies that are usually hidden in analyses restricted to the local rest frame, by automatically excluding theories of hydrodynamics that are not covariantly stable. 
Because the foliation of spacetime is arbitrary, we have found that the equilibrium distribution for fluctuations allows for the calculation of correlations not only at equal time but for any points at mutually spacelike separations.

To demonstrate the generality of our formalism, and to illustrate issues connecting causality and thermodynamic stability, we have presented applications to three different theories. First, we have shown how causality issues in relativistic Navier-Stokes theory in the Landau frame lead to unstable fluctuations, and how these issues are masked in the local rest frame. We also applied our approach to a relativistic model for heat diffusion equation a la Cattaneo, carefully analyzing the limit of vanishing relaxation time, in which the theory becomes acausal and unstable. 

Finally, we have applied our framework to the case of conformal Israel-Stewart hydrodynamics in a general hydrodynamic frame, that is, in the presence of an energy diffusion current and an off-equilibrium correction to the energy density.  After introducing arbitrary noise sources, we have determined that stochastic noise terms should be introduced exclusively in the equations of motion for the off-equilibrium corrections. We computed the corresponding noise correlators and symmetrized two-point functions.  We also found that, while IS theory in the Landau frame is able to produce the most general form for the one-point function of the energy-momentum tensor, the corresponding theory in a general frame leads to a more general form for the symmetrized two-point function. This suggests that taking the Landau frame artificially removes fluctuating modes, leading to a loss of generality. That finding motivates the further investigation of fluctuations in second-order theories in a general hydrodynamic frame.

The next step in the development of the present formalism would be its extension beyond second-order in fluctuations, which would enable the investigation of higher-order moments and cumulants of fluctuations. 
Most importantly, such an extension would allow us to explore the renormalization of transport coefficients \cite{Kovtun:2011np,Kovtun:2012rj,Abbasi:2022aao}.

A pressing outlook is the application of our results to the study of relativistic hydrodynamics near a critical point \cite{An:2021wof}. Such a study could be relevant for the ongoing search for the QCD critical point, and the techniques developed in this paper allow for fluctuations caused by critical phenomena to be included in a wide variety of models in a causal and stable manner. Furthermore, it would be interesting to investigate how the interplay between dissipation and noise in the new type of universality classes in relativistic hydrodynamics discussed in \cite{Gavassino:2023odx,Gavassino:2023qwl}.

Finally, in a companion paper we will discuss how to formulate the approach presented here from an action principle, and compare it with standard effective field theory approaches to hydrodynamics (e.g. \cite{Liu:2018kfw}).

\section*{Acknowledgements}

We thank L.~Gavassino for discussions about the information current of Navier-Stokes theory, and for insightful comments about the manuscript concerning the fate of fluctuations and dissipation in  relativistic fluids. 
NM and JN are supported in part by the U.S. Department of Energy, Office of Science, Office for Nuclear Physics
under Award No. DE-SC0021301. MH and JN were supported in part by the National Science Foundation (NSF) within the framework of the MUSES collaboration, under grant number OAC-2103680. JN thanks KITP Santa Barbara for its hospitality during ``The Many Faces of Relativistic Fluid Dynamics" Program, where this work's last stages were completed. This research was supported in part by the National Science Foundation under Grant No. NSF PHY-1748958.

\bibliography{references,References_Jorge}

\end{document}